\documentclass[aps,prl,twocolumn,superscriptaddress,showpacs,amsmath,amssymb,longbibliography]{revtex4-1}
\usepackage{graphicx}
\usepackage[hypertexnames=false,breaklinks]{hyperref}

\usepackage[utf8]{inputenc}
\usepackage{polski}
\usepackage[english]{babel}
\usepackage{amsmath}
\usepackage{xcolor}
\usepackage{ulem}

\makeatletter
\newcommand*{\balancecolsandclearpage}{%
  \close@column@grid
  \clearpage
  \twocolumngrid
}

\usepackage[sort&compress]{natbib}
 
\begin{document}

\title{Identification of Majorana Modes in Interacting Systems by Local Integrals of Motion.}%
\author{Andrzej Wi\k{e}ckowski}%
\author{Maciej M. Ma\'ska}
\affiliation{Institute of  Physics,  University  of  Silesia,  40-007  Katowice,  Poland}
\author{Marcin Mierzejewski}
\affiliation{Department of Theoretical Physics, Wroc\l aw University of Science and Technology, 50-370 Wroc\l aw, Poland}

\begin{abstract}
Recently, there has been substantial progress in methods of identifying  local integrals of motion in interacting integrable models or in 
systems with many-body localization. We show that one of these approaches can be utilized  for constructing local, conserved, Majorana fermions in systems with an arbitrary many-body interaction. As a test case, we first investigate a non-interacting Kitaev model and demonstrate  that this approach perfectly reproduces the standard results.   Then, we discuss how the many-body interactions influence the spatial structure and the lifetime of the Majorana modes. Finally, we determine the regime for which the information stored in the Majorana correlators is  also  retained  for arbitrarily  long  times  at high  temperatures.
We show that it is included in the regime with topologically protected soft Majorana modes, but in some cases is significantly smaller. 
\end{abstract}

\maketitle

{\it Introduction.---}Recently, a lot of hope has been pinned on Majorana zero modes as building blocks of a quantum computer \cite{QC1,QC2,QC3,QC4,Akhmerov}. 
%
%
One of the systems where these modes were proposed and observed is a semiconductor nanowire with a spin-orbit interaction coupled to an $s$-wave superconductor \cite{PhysRevLett.105.077001, PhysRevLett.105.177002, Franz2013, Mourik1003,Nadj-Perge602,Pawlak2016,Ruby2015}.
%
It is known that, in low-dimensional systems, Coulomb interactions are crucial and can drastically affect their properties \cite{Haldane,Gangadharaiah,superconductor_driven,envir,intdod}.
Interactions are also important  for practical reasons:
disorder is present in any semiconductor nanowire and the Majorana states are not completely immune against it \cite{Maska,Lutchyn2011,Akhmerov2011}. 
Moderate interactions may stabilize  the Majorana states against such perturbations \cite{pinning,Stoudenmire,Gergs,Hassler2012}.

Generally, a Majorana fermion is any fermionic operator $\Gamma$ that satisfies $\Gamma^2=1$. 
However, in order to perform topological quantum computing one needs
stable, non-Abelian anions \cite{RevModPhys.80.1083}. They can be realized as 
localized  {\it Majorana zero modes} (MZMs), whereby their stability follows from the  commutation relation 
\begin{equation}
  \left[\hat{H},\,\Gamma\right]=0,
  \label{comutator}
\end{equation}
where $\hat{H}$ is the Hamiltonian. 
This equation together with the conservation of the fermion parity lead to a non-Abelian braiding for adiabatic exchanging of  Majorana quasiparticles \cite{nonabelian}.
Equation~(\ref{comutator}) can be fulfilled  rigorously only in the thermodynamic limit, except for a fine-tuned symmetric point   \cite{kitaev} where it also holds true  for finite $L$. 
 In general,  $\left[\hat{H},\,\Gamma\right]\propto e^{-L/\xi}$,  where $L$ is the  system size and $\xi$ is correlation length \cite{Kells}.
The nonzero value of the commutator means that, even in the absence of any external decoherence processes, the MZM will have a finite lifetime.

The question is how to find the topological order and Majorana modes in interacting systems.  Several methods  have been used to study MZMs in interacting nanowires  \cite{Ng,Su,Chan,Hofmann,thomale}, see Ref. \cite{Stoudenmire} for a review.  A commonly tested  {\it necessary} condition [which follows from Eq. (\ref{comutator})] concerns degeneracy of the ground states obtained for systems with odd and even numbers of fermions.  A {\it sufficient} condition for the presence of topological order  is more involved.  It can be formulated  based on the local unitary  equivalence (LUE)  between the ground states  of the interacting system and of the noninteracting Kitaev chain in the topological phase \cite{class2}. In order to prove LUE, it is sufficient to
show that one of the ground states  can be continuously deformed to the other, whereby the spectral gap above the ground state must stay open along the entire path of deformation \cite{class1,Katsura}. 
But this is not equivalent to Eq.~(\ref{comutator}) and guarantees only the so-called {\it soft mode}, which is fully protected by topology only at temperatures well below 
 the spectral gap. In other words,  a soft MZM commutes with the Hamiltonian which is projected into a low-energy subspace \cite{StrongZeroMode1}.
 At higher temperatures, the information encoded in this mode can  be lost after some time.


In this Letter, we propose a method that allows one to find Majorana operator $\Gamma$ that {\it almost} satisfies Eq.~(\ref{comutator})
within the entire Hilbert space. Our method finds the so-called {\it strong} MZM
that is stable at arbitrary high temperatures \cite{prethermal,Goldstein,Kemp2017}.  Perturbative construction of almost strong MZMs has recently been reported in Ref. \cite{Kemp2017}
for  the Ising--like model with nearest- and (integrability-breaking) next-nearest-neighbor interactions.  In contradistinction, our approach is general and can be applied for arbitrary Hamiltonians, in principle, also, for spinful fermions. 
To this end, we derive the optimal form of a {\it local} operator $\Gamma$ that guarantees the longest lifetime of the MZM. We determine the regime of existence of a strong MZM
and show that it is smaller than the regime with soft modes, the latter being established from LUE. 




{\it The general method.---}We consider the Hamiltonian $\hat{H}=\sum_m  E_m | m \rangle \langle m| $ and assume that the relevant degrees of freedom can be expressed in terms of the standard fermionic operators 
$a_j$ and $a^{\dagger}_j$ or, equivalently,  in terms of the Majorana fermions  $\gamma_{2j}= a_j+a_j^\dagger$,  $\gamma_{2j+1} = i( a_j^\dagger- a_j)$. Here, $j$ includes all quantum numbers, e.g., the spin projection. 
We search for  particular combinations of the Majorana operators 
$\Gamma=\sum_i \alpha_i  \gamma_i$ with real coefficients $\alpha_i$ such that $\Gamma$  is conserved \cite{multigamma}. We assume normalization $\sum_i\alpha_i^2=1$ when $\Gamma^2=1$. 
The conservation of $\Gamma$ can  conveniently be  studied  by  averaging this quantity over an infinite time window
\begin{eqnarray}
\bar{\Gamma}&=&\lim_{\tau' \rightarrow \infty} \frac{1}{\tau'}\int_0^{\tau'} {\mathrm d}t e^{iHt} \Gamma e^{-iHt},  \label{ta1} \\
&=& \lim_{\tau \rightarrow \infty}  \sum_{m,n} \theta\left(\frac{1}{\tau} -|E_m-E_n| \right) \langle n |\Gamma |m \rangle \;\;  |n\rangle \langle m |  \label{ta}.
\end{eqnarray}
If this mode is strictly conserved then $\bar{\Gamma}=\Gamma$.
This, however, would require Eq.~(\ref{comutator}) to be satisfied, what
may not be the case in finite systems. Therefore, 
 we will usually search for an optimal choice of  $\alpha_i$ when $\bar{\Gamma}$ is as {\it close } to $\Gamma$ as possible.  
In order to quantify the  proximity of two operators we use the usual (Hilbert-Schmidt) inner product $\langle  \hat{A} \hat{B} \rangle= \mathrm{Tr}(  \hat{A} \hat{B})/\mathrm{Tr}(\hat{1}) $.
The optimal choice of coefficients $\alpha_i$  corresponds to a minimum  of 
$ \langle  (\Gamma-\bar{\Gamma})^2 \rangle =   1-  \langle \bar{\Gamma}^2 \rangle$. 
The latter equality originates from the identity   $\langle  \Gamma \bar{\Gamma}\rangle = \langle   \bar{\Gamma} \bar{\Gamma}\rangle$
(i.e., the time averaging is an orthogonal projection), 
as shown in the Supplemental Material \cite{supp}. 
Consequently, the least decaying mode can be found from the optimization problem
\begin{equation} 
\lambda=\max_{\{ \alpha_i \}}\,  \langle \bar{\Gamma}^2 \rangle=\max_{\{ \alpha_i \}} \, \langle\bar{\Gamma}  \Gamma\rangle. \label{ldef}
\end{equation}   
 
 The physical meaning of $\lambda$ comes from  the observation that the scalar product $\langle ... \rangle $ formally represents  thermal averaging carried out for infinite temperatures. Then, following Eq. (\ref{ldef}), $\lambda$  is the asymptotic value of the longest living autocorrelation  function $\langle \Gamma(t) \Gamma \rangle $.  If  $\lambda=1$, then $\Gamma$ is a strict integral of motion. i.e. a strong MZM \cite{prethermal,StrongZeroMode0,StrongZeroMode1,StrongZeroMode2}.  For $0<\lambda<1$ the information stored in the correlator 
$\langle \Gamma(t) \Gamma \rangle $ is partially retained for arbitrarily long times (despite $\Gamma$ not being strictly conserved), while this information is completely lost when  $\lambda=0$. 
The optimization problem can be further simplified
\begin{equation} 
\lambda=\max_{\{ \alpha_i \}}   \sum_{ij} \alpha_i \langle \bar{\gamma}_i  \bar{\gamma}_j  \rangle  \alpha_j  \label{ldef1}.
\end{equation}    
 It becomes a standard eigenproblem for the (positive semidefinite) matrix $\langle \bar{\gamma}_i  \bar{\gamma}_j  \rangle$. Namely, $\lambda$ is the largest eigenvalue of $\langle \bar{\gamma}_i  \bar{\gamma}_j  \rangle$, and $\alpha_j$ are components of the corresponding eigenvector. 
Essentially, all nonvanishing eigenvalues (whether degenerate or not) correspond to  {\it independent} MZMs, whereby  their independence follows from 
orthogonality of different eigenvectors and the identity $\langle \gamma_i \gamma_j \rangle=\delta_{ij} $.

The general idea behind  this method is similar to another approach  which has previously been used for identification of new integrals of motion in the Heisenberg model \cite{mm1}.
The latter approach targets operators which are conserved and local. Here,
we single out  Majorana operators which are conserved and, at the same time, are local. 
The conservation follows from the time averaging, i.e.,  from the identity $[\hat{H},\bar{\Gamma}]=0$, whereas locality originates from the fact that  $\Gamma$  is a linear
combination of $\gamma_i$, each of them being supported on a single site only. Since we maximize the projection $\langle \bar{\Gamma} \Gamma \rangle $, the resulting operators 
retain the properties of both $\Gamma$ and $\bar{\Gamma}$; i.e., they are local, conserved MZMs.  More formal discussion concerning MZMs  (including their locality \cite{OBrien:2015aks}) can be found in the Supplemental Material \cite{supp}. 


 When studying systems with fixed boundary conditions, it  is utterly important,  that the limit for the size of the system $L \rightarrow \infty$ precedes  the limit for time $\tau \rightarrow \infty $,  \cite{shastry,sirker2014}.  Since numerical calculations can be  carried out for finite systems only, $\tau$  in Eq. (\ref{ta}) should be kept large but finite until the finite--size scaling 
is accomplished. All the discussed properties of the correlation functions also hold true  for finite  $\tau$ \cite{mm2,annalen}, even though it is not the case for finite  $\tau'$ in   Eq. (\ref{ta1}).

{\it Example.---}As an example, we study a one--dimensional  system of interacting, spinless fermions with hard-wall boundary conditions.  The system is described  by the Kitaev Hamiltonian \cite{kitaev} extended by the many body interactions
 
\begin{eqnarray}\hat H &=& -t_0\sum_{i=1}^{L-1}  \left(   a_{i+1}^\dagger a_{i}+\text{H.c.}\right)
+\Delta \sum_{i=1}^{L-1}\left(a_{i+1}^\dagger a_{i}^\dagger  + \text{H.c.} \right)\nonumber\\
&& -\mu\sum_{i=1}^{L}\widetilde n_i 
+ V\sum_{i=1}^{L-1} \widetilde n_i \widetilde n_{i+1}
+W\sum_{i=1}^{L-1}\widetilde n_i \widetilde n_{i+2}. \label{ham}
\end{eqnarray}
Here, $t_0$ refers to hopping amplitude, $\mu$ is a chemical potential,
$\Delta$ is the superconducting gap  and  $\widetilde n_i= a_i^\dagger a_i-\frac{1}{2}$.  $V$ and $W$ are potentials of the first and second nearest-neighbor interactions. 
 For simplicity, we use  dimensionless units by putting $\hbar=1$ and $t_0$=$1$.

\begin{figure}[h!]
\centering
\includegraphics[width=\columnwidth]{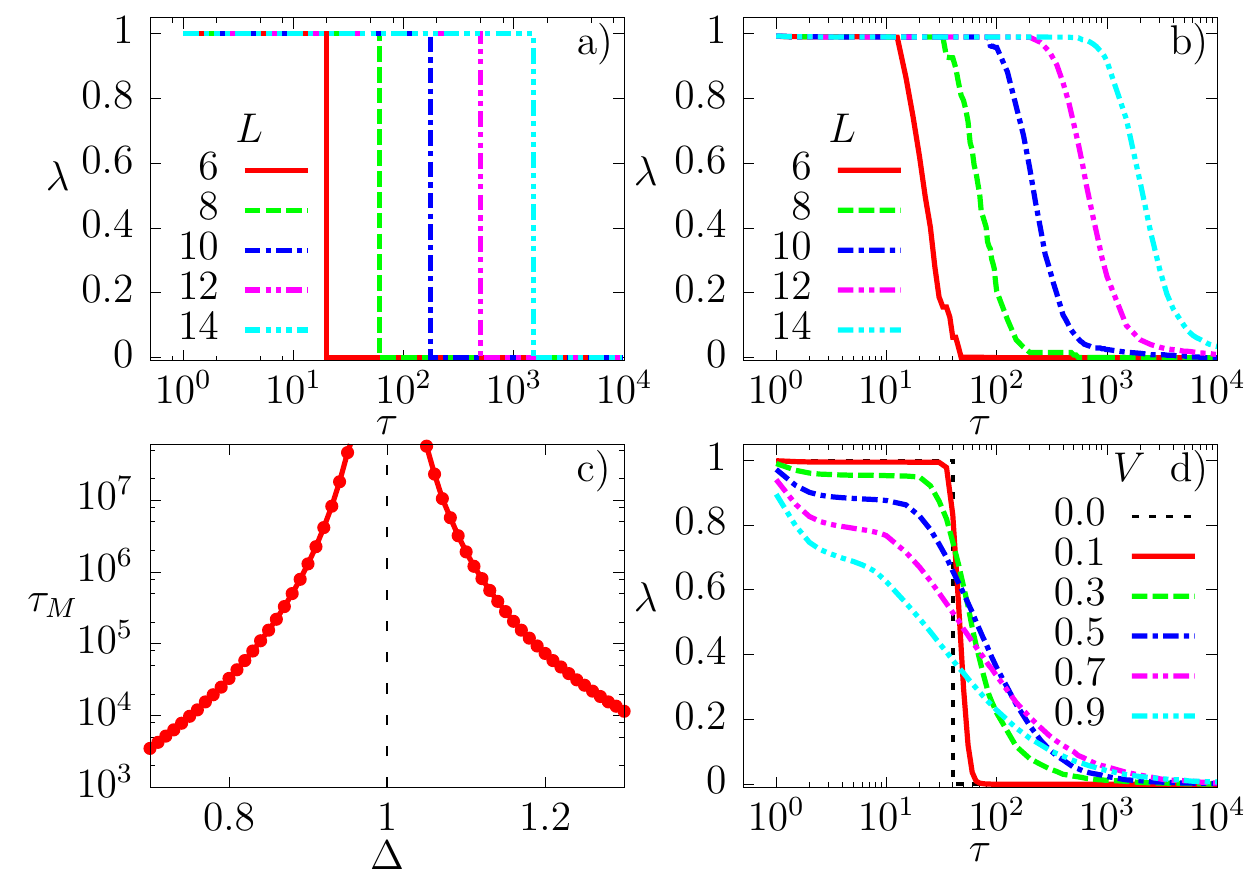}
\caption{ Results for systems without (a,c) and with (b,d) many-body interactions and $\mu=0$. 
 (a), (b) and (d) The Majorana autocorrelation function $\lambda$ [see Eq. (\ref{ldef})] for: (a) $V=0, \Delta=0.5$; (b) $V=0.2, \Delta=0.5$; (d) $L=12,\Delta=0.3$. (c) Lifetime of MZMs  for a finite noninteracting system of $L=10$ sites.}
\label{fig1}
\end{figure} 

\begin{figure}
\centering
\includegraphics[width=\columnwidth]{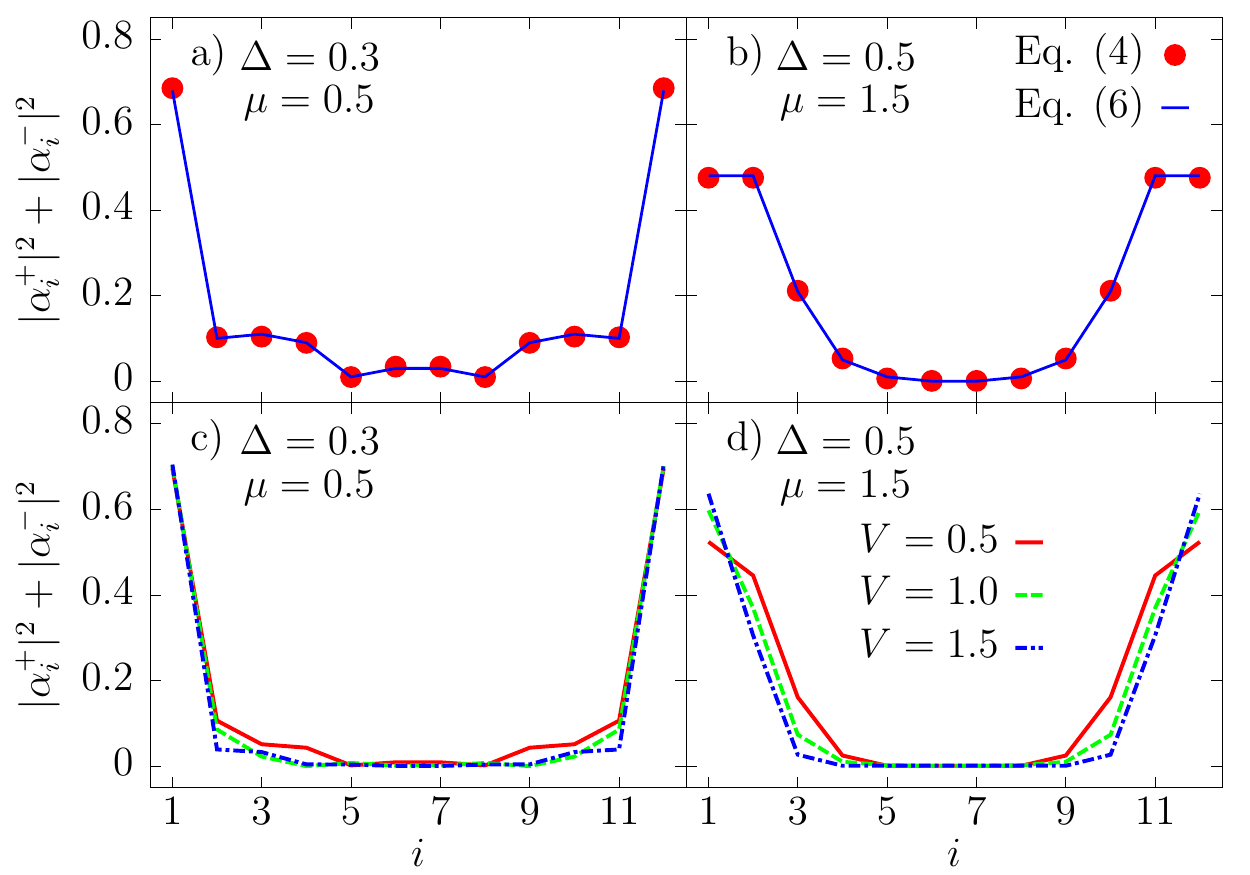}
\caption{
Spatial  structure of MZMs, $\Gamma^+=\sum_i \alpha^+_i \gamma_{2i}$ 
and  $\Gamma^-=\sum_i \alpha^-_i \gamma_{2i+1}$.  a) and b) Rescaled local density of states at energy $E=0$ for noninteracting system [$V=0$ , Eq.~(\ref{nonint})] (solid line)
compared with solution of Eq. (\ref{ldef1}) (points). c) and d) Results for $V \ne 0$ from Eq. (\ref{ldef1}).
}
\label{fig2}
\end{figure} 

{\it Test for noninteracting systems.---}Numerical implementation of our approach consists of three consecutive steps: {\it (i)} exact  diagonalization of the Hamiltonian (\ref{ham});  {\it (ii)}  numerical construction of time--averaged
Majorana operators  $\bar{\gamma}_i$  as defined by Eq. (\ref{ta}) but for finite $\tau$;  {\it (iii)}  construction and diagonalization of the matrix $K_{ij}=\langle \bar{\gamma}_i  \bar{\gamma}_j  \rangle$.
Because of the orthogonality relation $\langle \bar{\gamma}_{2i} \bar{\gamma}_{2j+1} \rangle=0$, one may separately study two cases $\Gamma^+=\sum_i \alpha^+_i \gamma_{2i}$ 
and  $\Gamma^-=\sum_i \alpha^-_i \gamma_{2i+1}$, whereby, now, the index $i$ enumerates the lattice sites. In the rest of this work, we discuss the two most stable modes (one in each sector  $\Gamma^+$ and $\Gamma^-$). All other eigenvalues of the matrix $K$ are much smaller and vanish in the thermodynamic limit (not shown).   It remains in agreement with a common knowledge that the homogeneous chain described by the Hamiltonian (\ref{ham}) may host at most two MZMs exponentially localized at the boundaries  \cite{kitaev,Kells,Katsura} . 

The complexity  of our approach  is independent of whether or not the many-body interactions are present:  hence, the method can be  tested by investigating a noninteracting system with  $V=W=0$.  
Figiure \ref{fig1}a) shows $\tau$ dependence of  $\lambda$ [see Eq. (\ref{ldef})]  for the most stable MZM  $\Gamma^+$.   Results for $\Gamma^-$ are exactly the same.     
One may introduce the lifetime of the MZMs, $\tau_{M}$, corresponding to the vertical sections of curves shown in the latter plot. 
Figure  \ref{fig1}c) shows that, for finite system, $\tau_{M}$ is finite as well, despite the absence of the many-body scattering.  The only exception concerns $|\Delta|=1$ when $\tau_{M} \rightarrow \infty$ for arbitrary  $L$. Otherwise, $\tau_{M}$  increases exponentially with $L$, as follows from the equal  spacing of the vertical sections in Fig. \ref{fig1}a).   The latter result clearly illustrates  the importance
of the correct order of limits: $\lim_{\tau \rightarrow \infty} \lim_{L \rightarrow \infty}  \lambda=1$; i.e., the MZMs  are strictly conserved in the thermodynamic limit,  
while  $ \lim_{L \rightarrow \infty}  \lim_{\tau \rightarrow \infty} \lambda=0 $.  All the obtained results remain in agreement with the well  established properties of the MZMs  in a noninteracting case, see,  e.g., \cite{Kells}. 

We have also calculated the local density of states at zero energy for the noninteracting Hamiltonian
\begin{equation}
  \rho_i(E=0)=-\frac{1}{\pi}{\rm Im}\,G_{ii}(E=0),\ \ \ G(E)=(E-\hat{H}+i\eta)^{-1},
  \label{nonint}
\end{equation}
where $\hat{H}$ is given by Eq.~(\ref{ham})  but with $V=W=0$.
In Figs. \ref{fig2}a) and  \ref{fig2}b), rescaled $ \rho_i(E=0)$  is compared with the spatial density of the Majorana fermions contributing to both Majorana modes, $|\alpha^+_i |^2+ | \alpha^-_i |^2$.  Perfect agreement between both methods illustrates accuracy of the approach derived in this work.

\begin{figure}
\includegraphics[width=\columnwidth]{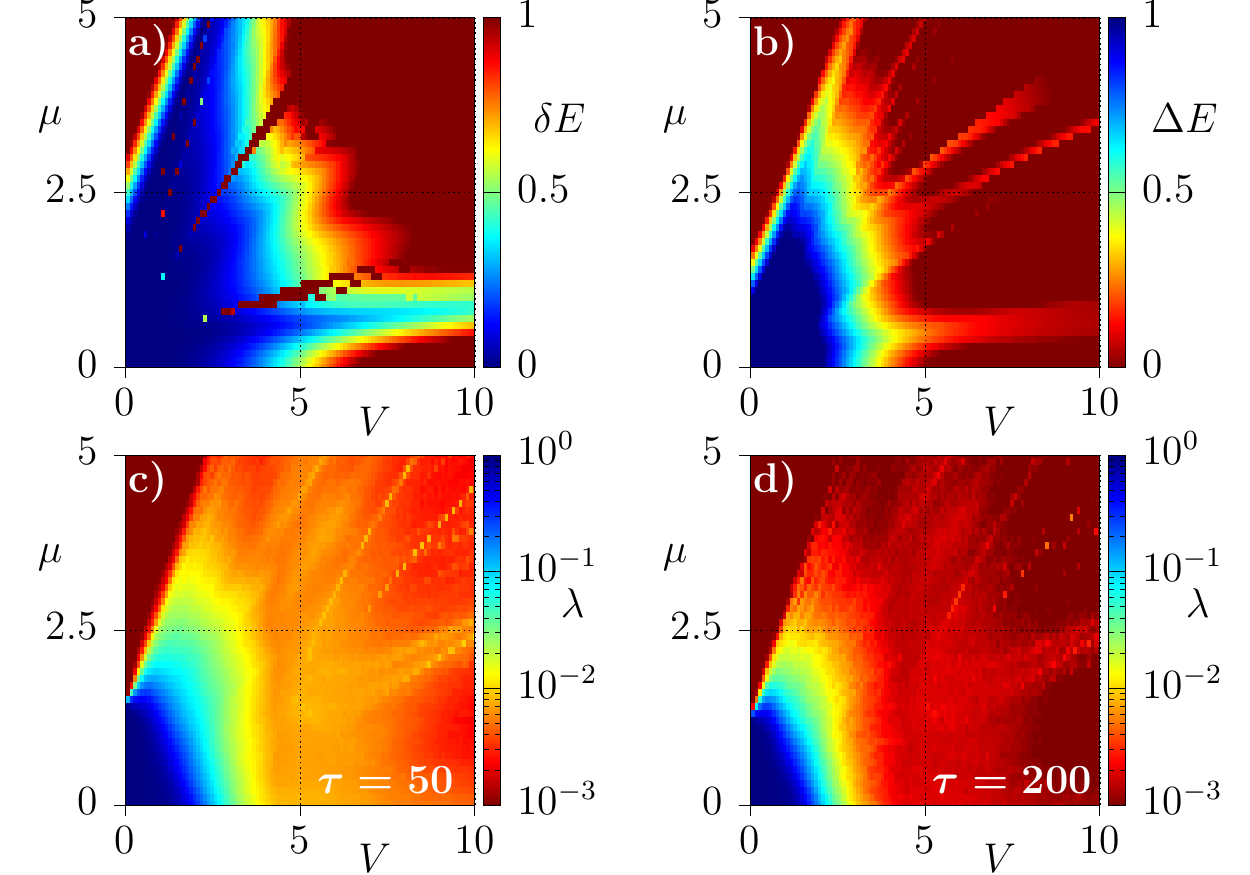}
\caption{Results for $\Delta = 1 $.  a) Degeneracy of the ground states. b) The spectral gap.  
c) and d) show the Majorana autocorrelation function $\lambda$ for various  times $\tau$ and $L=12$.
Note different color schemes in a) and b).
}
\label{fig3}
\end{figure} 
 
 \begin{figure}
\centering
\includegraphics[width=\columnwidth]{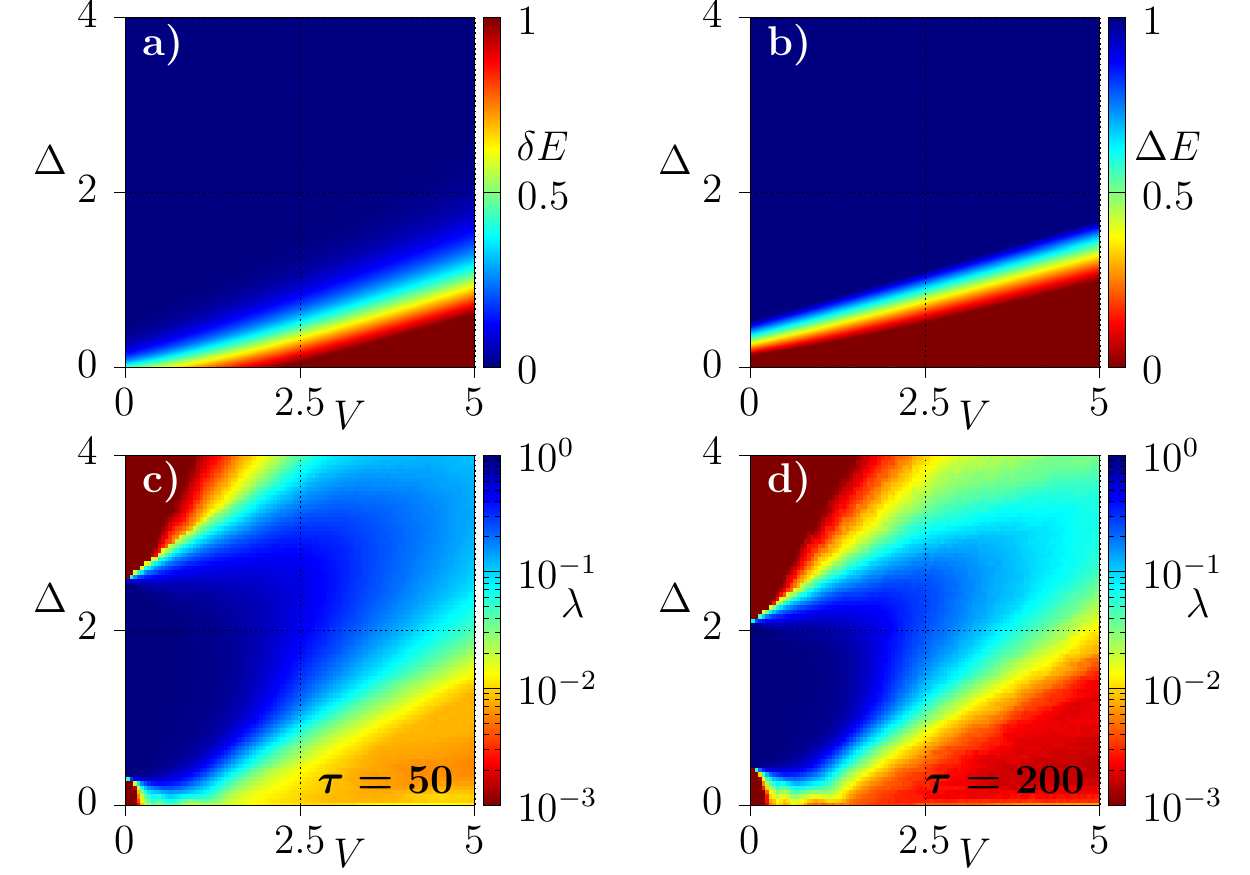}
\caption{The same as in Fig. \ref{fig3}, but as a function of $V$ and $\Delta$ for $\mu=0$.
}
\label{fig4}
\end{figure} 

 \begin{figure}
\centering
\includegraphics[width=\columnwidth]{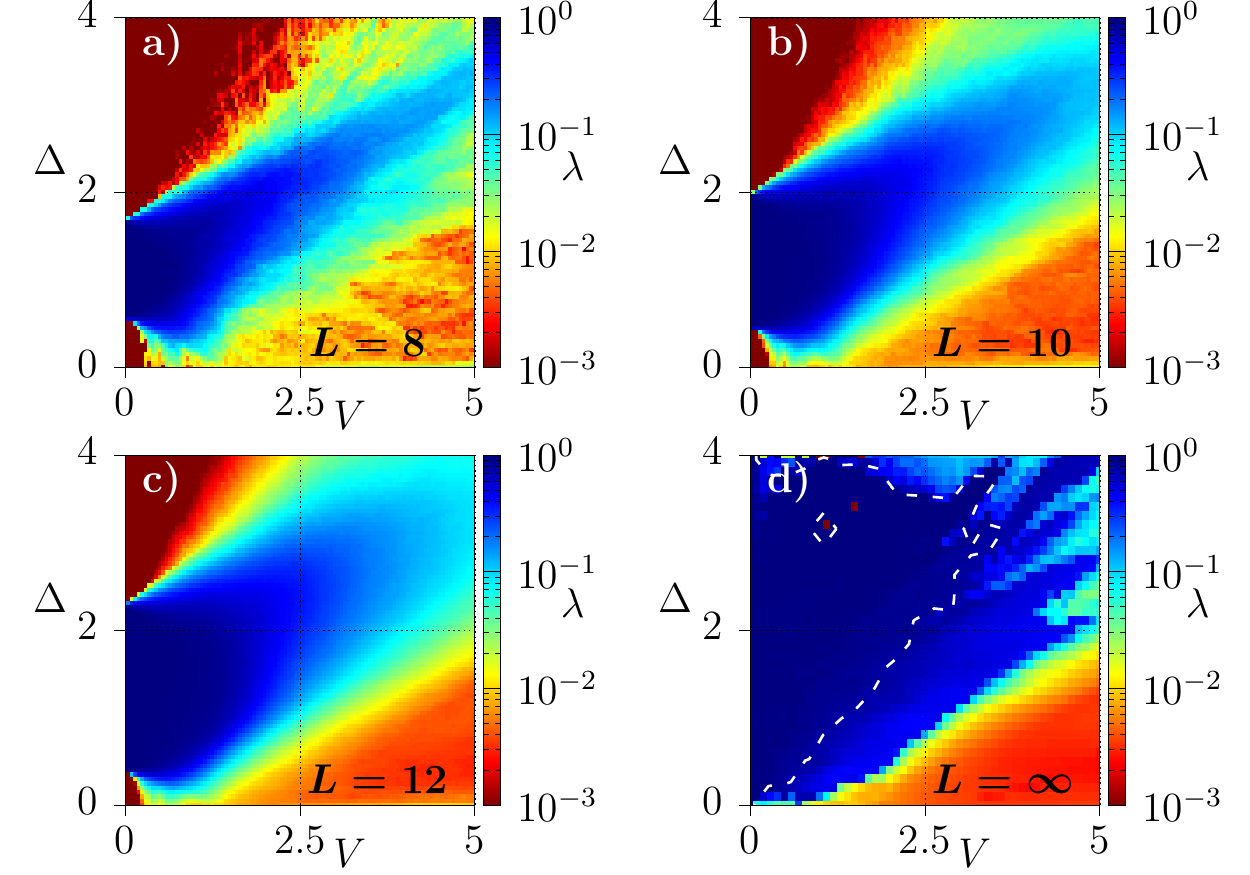}
\caption{
The autocorrelation function $\lambda$ as a function of $V$ and $\Delta$ for $\tau=100$ and various system sizes $L$.  Contour in d) marks $\lambda=0.8$.}
\label{fig5}
\end{figure}

{\it Systems with many-body interactions.---}All results in the main text  will be shown
for $W=V/2$, whereas the commonly studied case $W=0$ (which contains some peculiar features)  is discussed in the Supplemental Material \cite{supp}. 
Results in Figs. \ref{fig1}b) and \ref{fig1}d) show the most stable Majorana autocorrelation function [Eq. (\ref{ldef})] in the presence of weak to moderate interactions. Similar to the noninteracting case [Fig. \ref{fig1}a)], the position of the steep sections of $\lambda(\tau)$ increases exponentially  with the system size indicating that   $\lim_{\tau \rightarrow \infty} \lim_{L \rightarrow \infty}  \lambda  \simeq 1 $ but, in contrast to noninteracting systems, $\lambda < 1$. In the Supplemental Material \cite{supp}, we show that  the latter inequality seems to be generic for  systems with many-body interactions. It  implies that the strictly local operator $\Gamma$ is not a strict integral of motion. 
 Our approach singles out $\Gamma$ which contains the largest possible conserved part represented by $\lim_{\tau \rightarrow \infty} \bar{\Gamma}$. 


For finite systems, the many-body interactions  may extend the time scale in which the correlator $\langle \Gamma(t) \Gamma \rangle $ is large.  
Interestingly,  this extension can exceed 1 order of magnitude, as shown in Fig. \ref{fig1}d). 
Figures \ref{fig2}c) and \ref{fig2}d) explain the origin of this extension.
They show how the many-body interactions modify the spatial structure of the MZMs.  There are two modes which vanish exponentially outside of the edges of the system. Note that this
  property is not built into our algorithm but appears as a result which doesn't need to hold true for other geometry of the system.  Despite the exponential decay, these two modes still do overlap  and this overlap is responsible for a finite-lifetime of the MZMs in a noninteracting system with $L < \infty$.
Then, the many-body interactions push these modes further towards the edges of the system [see Figs.  \ref{fig2}c) and \ref{fig2}d)], reducing the overlap between them and, in this way, increasing their lifetime. This mechanism holds true as  long as the interactions are not too strong, when the MZMs  eventually disappear. 

Next, we compare our results for strong MZMs with the presence of the topological order. We check the degeneracy of the ground  state (necessary condition) as well as LUE to the topological regime in the noninteracting Kitaev model (sufficient condition).
 To this end, we study chains of $L=8, 10,..., 20$  and find the two lowest energies in the subspaces with odd and even particle numbers, denoted, respectively, as $E_{0,o}(L)$, $E_{1,o}(L)$ and  $E_{0,e}(L)$, $E_{1,e}(L)$. We introduce the  measure of the ground--state degeneracy
$\delta E (L)= E_{0,o}(L)- E_{0,e}(L)$  and two spectral gaps, $\Delta E_{o(e)}(L)=E_{1,o(e)}(L)-E_{0,o(e)}(L)$.
Typically, the  gaps between the low--energy levels decay algebraically with $L$; hence, we carry out linearly in $1/L$  extrapolations of   $\Delta E_{o(e)}(L)$.
However, $\delta E(L)$ should decay exponentially  in the topological
regime; thus, we use the fitting function $\delta E (L)=A \exp(-B L)+\delta E (\infty)$. These extrapolations break down when $V$ and $\mu$ are large \cite{supp}, what shows up as large errors for the extrapolated quantities, $\sigma_{\delta E}$ and  $\sigma_{\Delta E} $.   We identify the degenerate ground states as a regime  where both  $|\delta E (\infty)|$ and $\sigma _{\delta E}$ are small, defining $\delta E\equiv |\delta E (\infty)| + \sigma _{\delta E}\ll 1$ as the lower bound on the degenerate region.
The LUE implies that the gap $\min\{\Delta E_o (\infty),\Delta E_e (\infty) \}$  doesn't vanish along a path that reaches the topological regime for $V=0$, while $\sigma _{\Delta E}$ remains small. Then, we define 
the lower bound on the corresponding region by $\Delta E\equiv \min\{\Delta E_o (\infty),\Delta E_e (\infty) \}- \sigma _{\Delta E}>0$.
Results for $\delta E$ and $\Delta E$ are shown in  Figs. \ref{fig3}a), \ref{fig4}a)  and  \ref{fig3}b), \ref{fig4}b), respectively. 
The actual  topological  region may be larger than it follows from lower bounds shown in Figs. \ref{fig3}b) and \ref{fig4}b).

Results in Figs. \ref{fig3}c) and \ref{fig3}d)  show that the  strong MZMs,  indeed, exist  for very long times ($\tau > 200$) not only in the ground state but, essentially, in the entire energy spectrum.  
We also confirm that  a moderate many-body interaction extends the range of $\mu$ where soft and strong MZMs  are present \cite{Stoudenmire}.   

In Fig.  \ref{fig4},  we show similar results but for $\mu=0$ and various magnitudes of the superconducting gap $\Delta$. In this case an exact solution is known but only
for $\Delta=1$ and $W=0$ \cite{Katsura,Miao2017}. 
For large $\tau$ and  $\Delta \gg 1$, the strong MZMs
seem to be absent even for very weak many-body interaction.  However, it is a finite-size effect that, again, shows how important is the correct order of limits for time and the system size. 
Therefore, in Fig.  \ref{fig5} we set  $\tau=100$ and show the Majorana autocorrelation function for various values of $L$ together with results  extrapolated to $L\rightarrow \infty$.  The details
of extrapolation and results for $\lim_{\tau\rightarrow \infty}\lim_{L \rightarrow \infty}  \lambda(\tau)$ are shown in the Supplemental material \cite{supp}.  
The regime with $\lambda>0$ covers roughly the entire topological regime determined via  LUE to the single-particle Kitaev model [compare Figs. \ref{fig4}b) and \ref{fig5}d)]. However,  $\lambda$ gradually decreases with increasing interactions, and a strong MZM with large $\lambda$ exists within a much smaller regime, as shown e.g., by the contour  in Fig. \ref{fig5}d). 

{\it Conclusions.---}We have proposed an approach for finding local (strong or almost strong) MZMs 
which can be implemented for an arbitrary many-body interaction.  We have found that even at elevated temperatures, the lifetime of these modes is long enough so that  they may  be used effectively  to store the information.   The regime where the strong MZMs exist (as quantified by large $\lambda$ in our approach)  is included, but is smaller than the regime which is unitarily equivalent to the topological regime in the single-particle Kitaev model. It means that not all topological states are equally protected to be useful in, e.q., quantum computing. At finite temperatures, the systems with weak many-body interactions are preferable; however, these interactions may still be significant, when compared to other energy scales in the system.
Our results also suggest that in systems with many-body interactions the strictly local Majorana operators are not strict integrals of motion, however, their autocorrelation function remains large for arbitrarily long times. 



This work is supported by the National Science Centre, Poland via Projects No. 2016/23/B/ST3/00647 (A.W. and M.M.)  and No. DEC--2013/11/B/ST3/00824 (M.M.M.).

\balancecolsandclearpage
\title{Supplemental Material for: \\ Majorana modes in interacting systems identified by searching for  local integrals of motion}
\author{Andrzej Więckowski}%
\author{Maciej M. Ma\'ska}
\author{Marcin Mierzejewski}
\affiliation{Institute of  Physics,  University  of  Silesia,  40-007  Katowice,  Poland}
\part{\large Supplemental Material}
\maketitle
\setcounter{equation}{0}
\setcounter{figure}{0}
\setcounter{table}{0}
\setcounter{page}{1}
\makeatletter
\renewcommand{\theequation}{S\arabic{equation}}
\renewcommand{\thefigure}{S\arabic{figure}}

In the Supplemental Material we show that the Majorana fermions singled out by our approach are indeed the edge strong zero modes.    
Further on, we discuss  results for a system where the many--body interactions are restricted to the neighboring lattice sites only. Finally, we present the details of the finite--size scaling.

\section{Majorana edge zero mode}

 In this section, we closely follow the formal concept of strong edge zero mode described in Ref. \cite{Kemp2017,prethermal} and also in \cite{,Goldstein}.
Such mode is an operator that maps an eigenstate in one symmetry sector to a state in another sector with the same energy up to finite-size corrections.
Here, the term "strong" means, that this mapping holds true for all  eigenstates of Hamiltonian.  As clearly explained in Ref. \cite{Kemp2017}, it is a stronger condition than necessary for a topological order.
Interestingly, the strong  character of the Majorana zero modes has been recognized already in the pioneer paper by Kitaev \cite{kitaev}.
Below we demonstrate that for $\lambda=1$ [see Eq. (4) in the main text]  the Majorana operator, $\Gamma=\sum_i \alpha_i  \gamma_i$, is indeed a strong zero mode.

Using the commutation relations $\{\gamma_i,\gamma_j \}=2\delta_{ij}$ one finds 
\begin{equation}
\Gamma^2=\frac{1}{2} \sum_{ij}   \alpha_i  \alpha_j \{\gamma_i,\gamma_j \}=\sum_i \alpha^2_i=1. \label{gam2}
\end{equation}
We formally rewrite $\Gamma$ as well as $\bar{\Gamma}$  [see Eq. (2) in the main text]   in the basis of the eigenstates of  Hamiltonian
\begin{eqnarray}
\Gamma&=& \sum_{m,n}   \langle n |\Gamma |m \rangle \;\;  |n\rangle \langle m |,  \label{gam} \\
\bar{\Gamma}&=& \sum_{m,n:E_m=E_n}   \langle n |\Gamma |m \rangle \;\;  |n\rangle \langle m |. \label {gambar}
\end{eqnarray} 
$\bar{\Gamma}$  is strictly conserved, since it commutes with Hamiltonian,
\begin{equation}
[\bar{\Gamma},\hat{H}]= \sum_{m,n:E_m=E_n}   (E_m-E_n) \langle n |\Gamma |m \rangle \;\;  |n\rangle \langle m |=0.
\end{equation}
In the main text we have shown that the condition $\lambda=1$ it equivalent to  $\Gamma=\bar{\Gamma}$, hence for $\lambda=1$  one obtains also $\bar{\Gamma}^2=1$. Then, we see that  
\begin{eqnarray}
1=\langle n | \bar{\Gamma}^2 |n \rangle = \sum_{m:E_m=E_n} | \langle n |\Gamma |m \rangle|^2, \label{gambar2}
\end{eqnarray}
holds true for arbitrary eigenstate $|n \rangle$.  It is evident that the matrix elements  $\langle n |\Gamma |m \rangle $ are nonzero only for states  with opposite parity of the particle number. 
Assuming that the energy spectrum is at most doubly degenerate, we see that for each state $ | m \rangle  $ in one symmetry sector (e.g. with even number of electrons)  there exists a state $| n \rangle $ in the other
sector  (e.g. with odd number of electrons) such that $E_m=E_n$. Using Eqs. (\ref{gam}) and (\ref{gambar2}) we obtain 
\begin{equation}
\Gamma |m \rangle = \exp(i \phi_n) |n\rangle, \label{map}
\end{equation}
hence $\Gamma$ is indeed a strong zero mode. 

Equation (\ref{gambar}) might suggest that numerical diagonalization of the Hamiltonian is sufficient to single out the Majorana  strong zero modes.  The only problem seems to be sorting the energies and finding the matching pairs of  states $|n\rangle$ and $|m \rangle$ in different symmetry sectors but with equal energies. 
However, numerical diagonalization can be carried out for finite systems only when the energies of states with opposite parities differ by a finite-size corrections.
Due to these corrections, one would have to find the pairs of states such that the difference of their energies vanishes in the thermodynamic limit, $L \rightarrow \infty $.  Implementation of such  numerical procedure
is highly nontrivial, especially that  the typical distance between the consecutive energy levels decays exponentially with $L$ for arbitrary extensive Hamiltonian, i.e., independently of the presence
of  the Majorana modes. In our approach the problem of the finite size corrections to the energy levels is resolved simply by allowing for  finite $\tau$ in Eq. (2) in the main text.  The correct order of limits
$\lim_{\tau \rightarrow \infty} \lim_{L\rightarrow \infty} \lambda$ simply means that the strict degeneracy exists only in the thermodynamic limit. Even if the problem of not perfectly degenerate states
could be resolved by some other method, then one needs to carry out an independent test whether the Majorana mode, as given by Eq. (\ref{gambar}), is a local operator with support at the edges of the system.  

In our approach the largest eigenvalue $\lambda$ obtained from Eq. (4) in the main text allows to single out the Majorana strong zero mode $\Gamma=\sum_i \alpha_i  \gamma_i$.
 As demonstrated in Fig. 2 in the main text,  the coefficients  $\alpha_i$ decay outside of regions located at the edges of the chain, hence
$\Gamma$ is also the edge mode. It is important to stress that $\alpha_i$ and $\lambda$ are determined by a single eigenproblem. Consequently, the presence and properties of strong zero modes in the many-body Hamiltonian are fully specified by this eigenproblem.  We are not aware of any other numerical algorithm, which allows to single out strong edge zero modes in arbitrary Hamiltonian with many-body interactions.

\begin{figure}
\centering
\includegraphics[width=\columnwidth]{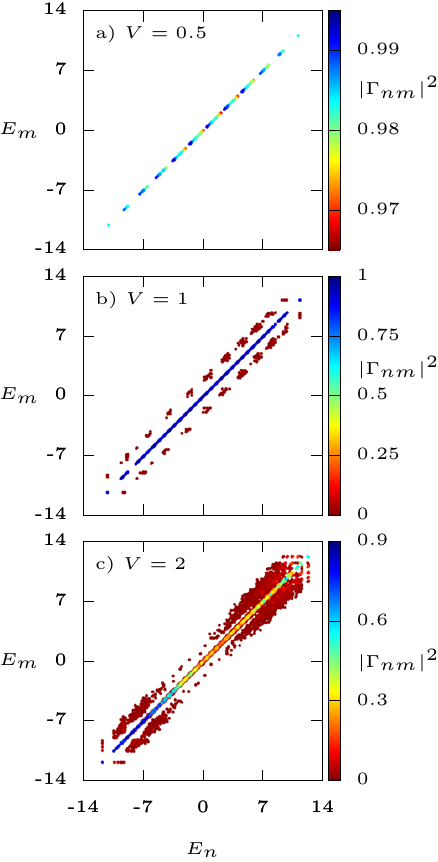}
\caption{
Squared matrix elements of the Majorana mode $|\Gamma_{nm}|^2= |\langle m |\Gamma | n \rangle|^2 $ for $L=12$ and $W=V/2$. Panels a,b and c show results for $V=0.5$, $V=1$ and $V=2$, respectively.  For such system we obtain at $\tau=10^2$, respectively, $\lambda\simeq 0.98$,  0.92 and 0.23. Note that each panel has a separate false color scheme.}
\label{S_dod}
\end{figure}

\section{Almost conserved Majorana modes}  
 
In this section we discuss the physical meaning of the Majorana fermions for $0 < \lambda < 1$. Using Eq. (3) in the main text, one obtains 
\begin{eqnarray}
\langle \Gamma \Gamma \rangle&=&\frac{1}{Z} \sum_{n,m} |\langle n |\Gamma | m \rangle|^2=1,  \label{eq1} \\
\langle \bar{\Gamma} \bar{\Gamma} \rangle&=&\frac{1}{Z}   \sum_{n,m:E_m=E_n} |\langle n |\Gamma | m \rangle|^2=\lambda \label{eq2}.
\end{eqnarray}
where $Z=\sum_n 1$ is the dimension of the Hilbert space. For $\lambda < 1$, $\Gamma$ remains a local operator (e.g., the edge mode), since it is defined as a linear combination
of the local Majorana operators $\gamma_i$.  However, it is not strictly conserved any more.
Using Eqs. (\ref{eq1}) and (\ref{eq2})  one finds for $\lambda <1$ that there are nonvanishing matrix elements $\langle n |\Gamma | m \rangle$
also for states with different energies $E_m \ne E_n$.  Therefore, for $\lambda < 1$ the strong character of the Majorana mode is lost.
Then, it is instructive to decompose the Majorana operator in the following way
\begin{eqnarray}
\Gamma= \bar{\Gamma} +\bar{\Gamma} ^{\perp},
\end{eqnarray}
where $\bar{\Gamma}$ is given by Eq. (\ref{gambar}). It represents the conserved part  (zero mode) of the Majorana operator, $[\bar{\Gamma},H]=0$. The remaining part
\begin{eqnarray} 
\bar{\Gamma} ^{\perp}= \sum_{m,n:E_m \ne E_n}   \langle n |\Gamma |m \rangle \;\;  |n\rangle \langle m |. \label {gambarper} 
\end{eqnarray}
is orthogonal to $\bar{\Gamma}$, $\langle \bar{\Gamma} \bar{\Gamma} ^{\perp} \rangle=0 $. Due to this orthogonality it is easy to find the (squared) norms of all operators:
\begin{eqnarray} 
||\Gamma ||^2 &=& \langle  \Gamma \Gamma \rangle=1, \\
||\bar{\Gamma} ||^2 &=& \langle  \bar{\Gamma} \bar{\Gamma} \rangle=\lambda,\\
||\bar{\Gamma}^{\perp} ||^2 &=& \langle  \Gamma \Gamma \rangle-\langle  \bar{\Gamma} \bar{\Gamma} \rangle=1-\lambda.
\end{eqnarray}
  
To conclude, our approach finds the Majorana fermions with the largest $\lambda$.  
It singles out the Majorana strong zero-mode if such mode exists ($\lambda=1$). Otherwise ($\lambda<1$), it finds the Majorana mode with the largest conserved part, $\bar{\Gamma}$.  

Figure \ref{S_dod}  shows the matrix elements of such Majorana mode $\Gamma_{mn}= \langle m |\Gamma | n \rangle $ as a function of energies $E_m$ and $E_n$. 
Fig.   \ref{S_dod}a shows results for $L=12, V=0.5$ and $W=V/2$. For such parameters one obtains $\lambda \simeq 0.98$  for $\tau=10^2$ hence $\Gamma$ is (almost) a strong Majorana mode. 
In agreement with Eq. (\ref{map}),  for each eigenstate $|n \rangle $ there exists a single state $|m \rangle$ with opposite parity  such that $|\Gamma_{mn}| \simeq 1$ and $E_m\simeq E_n$.
The middle panel shows results for the same system but with $V=1$ when $\lambda \simeq 0.92$  for $\tau=10^2$. Now, $\Gamma$ is not a strong zero mode any more. It contains a significant
 conserved part (zero mode), $\bar{\Gamma}$,  represented by points along the diagonal and much smaller not conserved part, $\bar{\Gamma}^{\perp}$, represented by the off--diagonal elements.  
 $\lambda$ is the ratio of contributions coming from the diagonal points [see Eq. (\ref{eq2})] to the total contribution coming from all the points  [see Eq. (\ref{eq1})].
 Finally, for even stronger interaction, $V=2$, we obtain $\lambda \simeq 0.23$ at   $\tau=10^2$ and the corresponding matrix elements are shown in Fig.   \ref{S_dod}c. 
 The conserved part is diminished mostly in the middle of the spectrum, but it still remains large at the bottom of the spectrum, what is relevant for the low--temperature regime. 
The  Hilbert-Schmidt inner product $\langle ... \rangle $ may be modified in such a way that it becomes relevant for finite temperatures, e.g. see Ref.  \cite{mm2}.  We have numerically studied selected cases also for $k_B T= 10 $  and found that the liftetime of the  MZMs is slightly larger than at $ T \rightarrow \infty$ (not shown).  Obviously, very low but nonzero temperatures are not accessible due to huge finite--size effects. 

\section{Results for a system with nearest neighbor interaction}

Numerical results presented in the main text have been obtained for repulsive interaction between the first ($V$) and  the second nearest neighbors  ($W$). 
Here, we discuss the same quantities but for the commonly studied case with $W=0$ which, however,  contains  some peculiar features.

\begin{figure}
\centering
\includegraphics[width=\columnwidth]{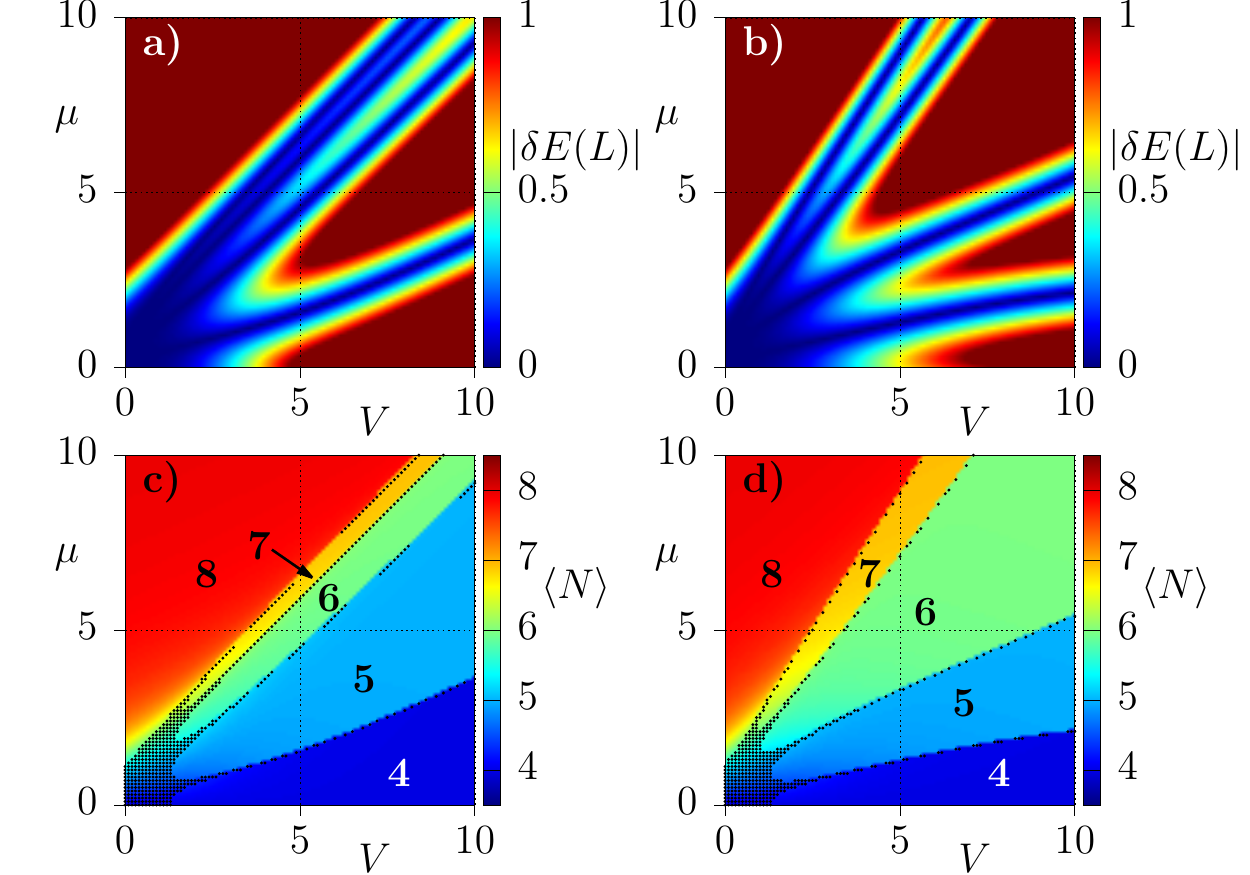}
\caption{
Results for  $L=8$, $\Delta=1$ and $W=0$ (a,c) or $W = V/2$ (b,d).
a) and  b)  Difference between the ground state energies obtained for different parities of fermions,  $\delta E(L)$. 
c) and d)  Average occupation of fermions in the ground state. Points mark parameters for  which $|\delta E(L)|<0.02$.
}
\label{S1}
\end{figure}

We start with the difference of the ground state energies $\delta E(L)$, obtained for systems with odd and even number of fermions.  Vanishing of  $\delta E$ in the
thermodynamic limit is {\it necessary} for the onset of soft Majorana modes. $\delta E(L) $ obtained for $L=8$  is shown in Figs. \ref{S1}a and \ref{S1}b for $W=0$ and $W=V/2$, respectively. The former figure accurately reproduces results presented in Ref. \cite{Ng}.  Here one may identify two regions where $\delta E(L)$ is small: (1) small 2D area around $V=0,\,\mu=0$ and (2) a number of narrow stripes that extend from region (1) to large--$V$ and large--$\mu$. While the internal structure of region (2)
is hardly visible for $W=0$, it becomes very clear for $W=V/2$, as shown in Fig. \ref{S1}b. It is composed of straight lines and their total number
 exactly equals the size of the system, whereby $L/2$ lines are explicitly visible in the figures, while remaining $L/2$ lines exist for  
$ \mu <  0 $.  The physical origin of these lines  is explained in Figs.  \ref{S1}c and \ref{S1}d where we show 
average particle number  in the ground state, $\langle N \rangle $. We also mark the parameters for which  $\delta E(L)$ is small. One can see that lines separate regimes where the ground state occupation is close to consecutive integers $0,1,\ldots,L$.  In the adjoining regimes, these integers have opposite parity, hence the borderline between these regimes corresponds to degenerate ground states  obtained in sectors with odd and even particle numbers.  However, such lines exist independently of the pairing term, i.e., also in topologically trivial phases, 
and are thus unrelated to the local Majorana modes.  The presence of the latter lines explains also why the finite--size scaling for $\delta E(L)$ breaks down in this regime.  
Since the number and positions of these lines change with $L$, $\delta E(L)$ may be non--monotonous functions of $L$. The same problem shows up also in the finite--size
scaling of the spectral gaps $\Delta E_{o}(L)$ and $\Delta E_{e}(L)$.

 \begin{figure}
\centering
\includegraphics[width=\columnwidth]{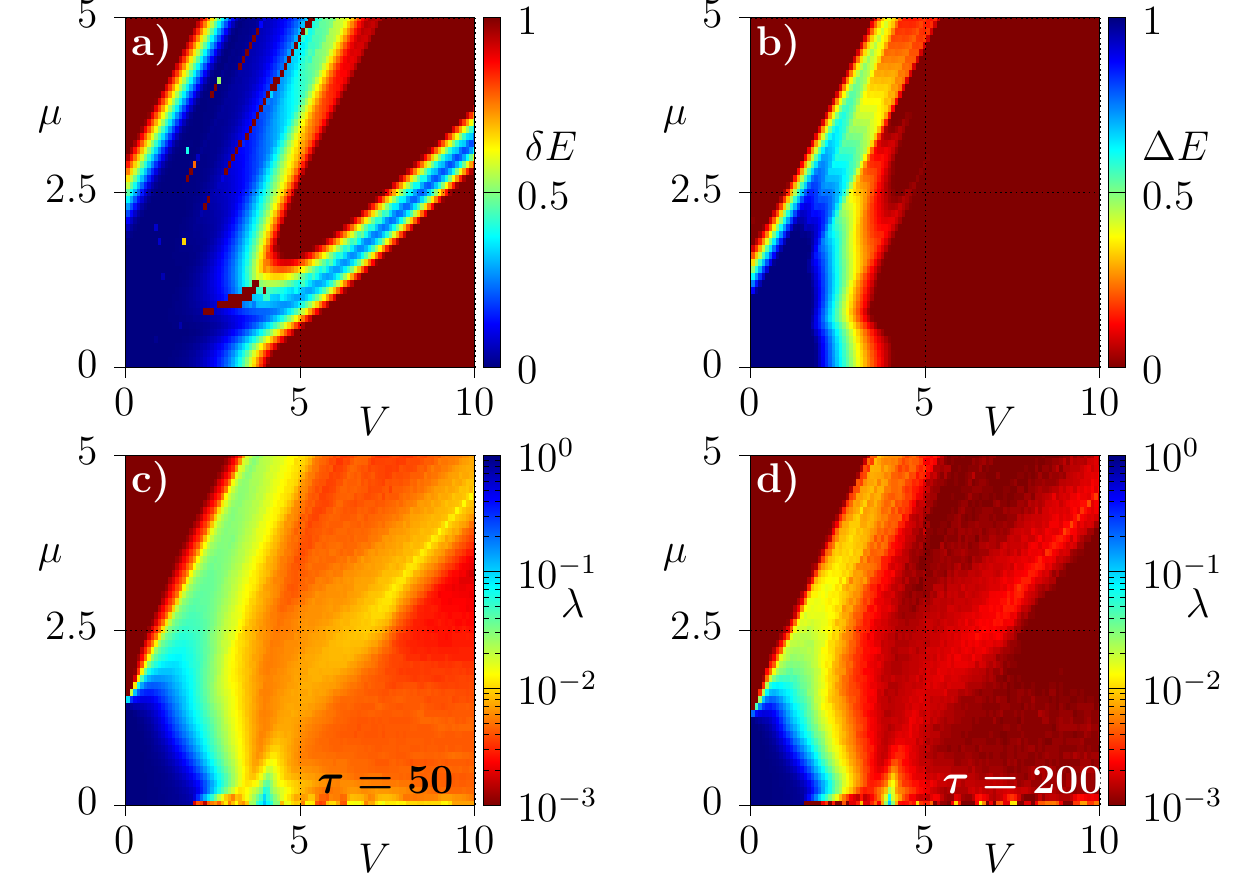}
\caption{
The same as in Fig. 3 in the main text, but for $W=0$.}
\label{S2}
\end{figure} 
 
 \begin{figure}
\centering
\includegraphics[width=\columnwidth]{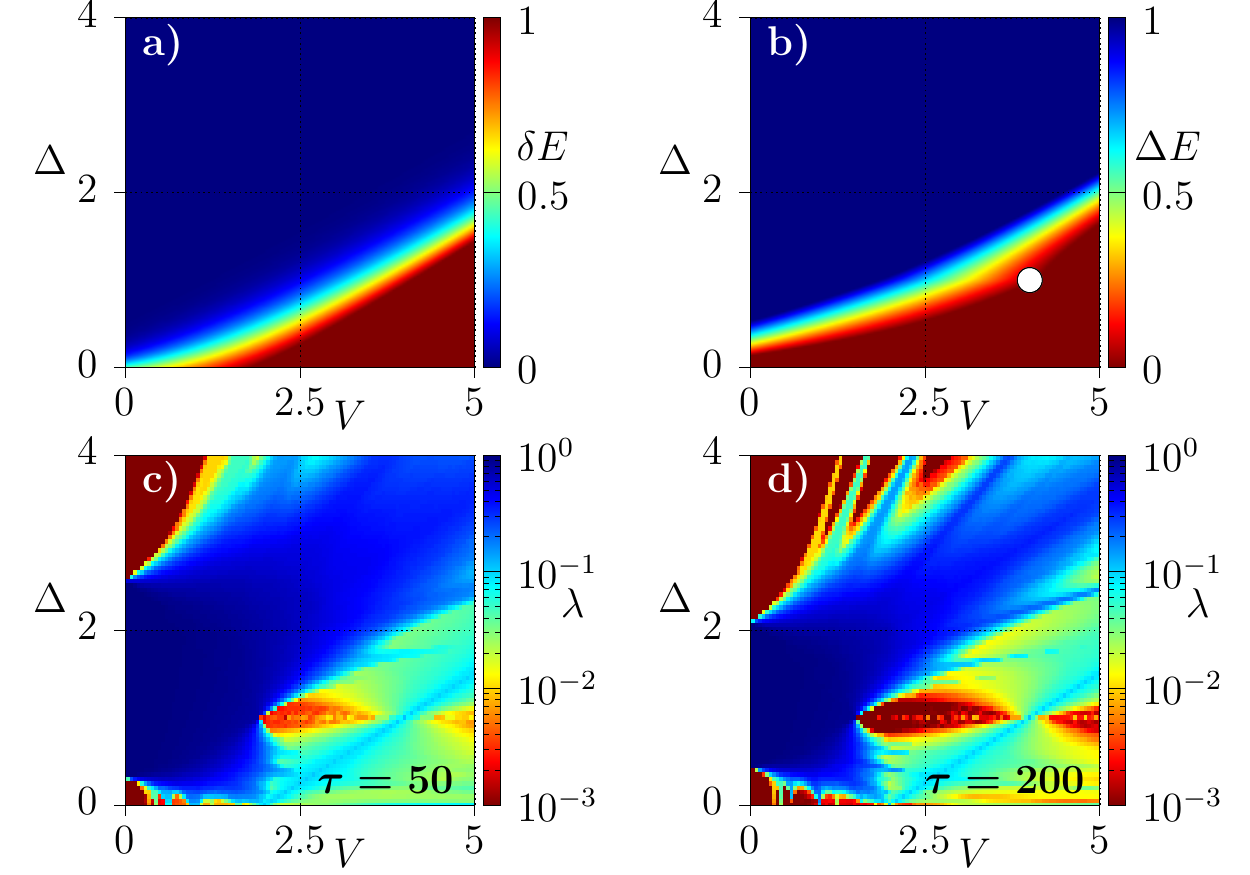}
\caption{
The same as  in Fig. 4 in the main text, but for $W=0$. Point in b) marks exact boundary of the topological phase from Ref. \cite{Katsura}.}
\label{S3}
\end{figure} 

\begin{figure}
\centering
\includegraphics[width=\columnwidth]{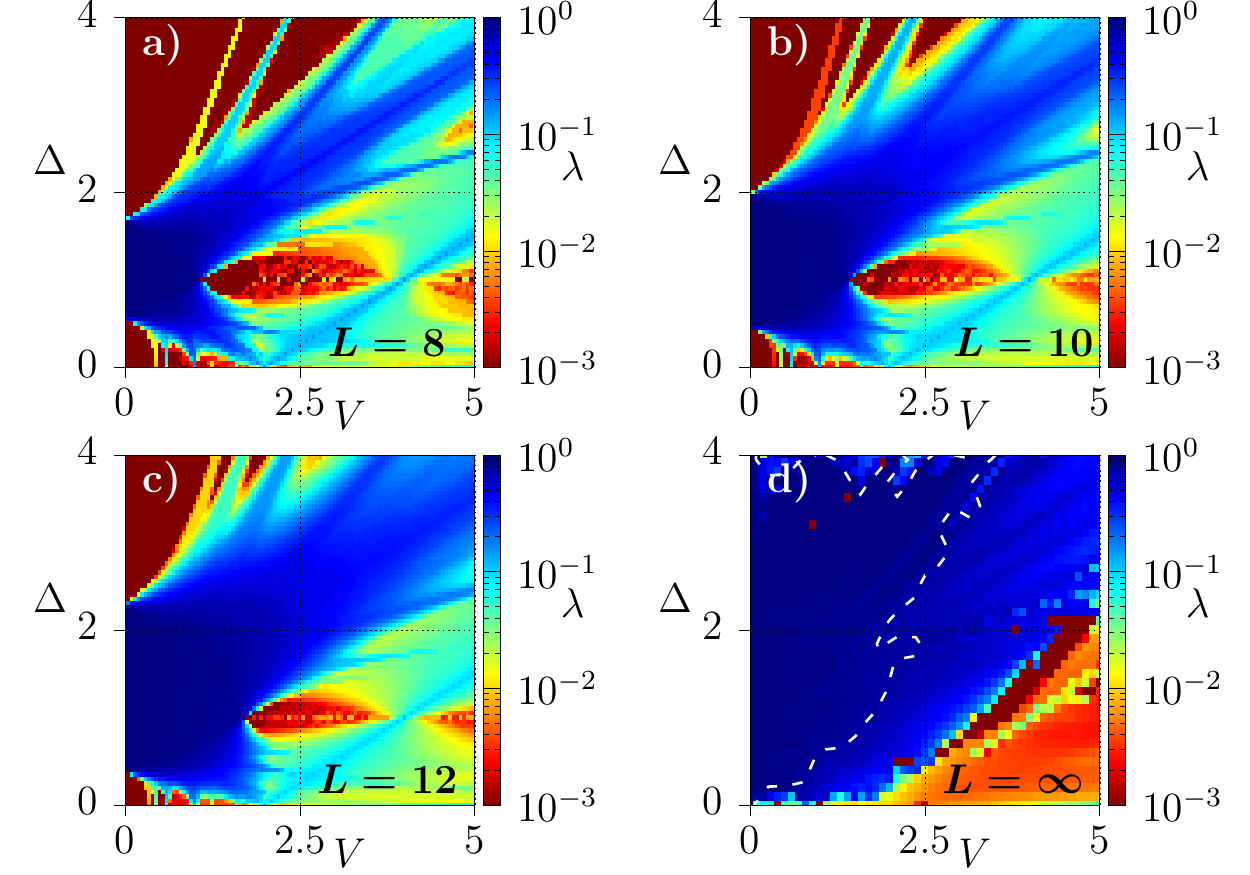}
\caption{
The same as Fig. 5 in the main text but for $W=0$.
 }
 \label{S4}
\end{figure}

Figures \ref{S2}, \ref{S3} and \ref{S4} show respectively the same phase diagrams as Figs. 3, 4 and 5 in the main text but for $W=0$. These phase diagrams are constructed from the Majorana autocorrelation function $\lambda$. On the one hand, these plots clearly show that the qualitative results discussed in the main text are generic, i.e., they are independent of a specific choice  of the many--body interaction. The main message is that the long--living Majorana modes exist not only in the ground state but within the whole energy spectrum for moderate interactions, while weak interaction may even expand the range of the chemical potential where these modes exist (see Figs. \ref{S2}c and \ref{S2}d).  On the other hand, the phase diagrams for $W=0$ include rather complicated structure of lines where 
Majorana modes are particularly robust, see e.g., Figs. \ref{S3}d and \ref{S4}a-\ref{S4}c.  However, such structures are not generic  because they don't show up for  $W\ne 0$.

\section{Details of the finite-size scaling}

As argued in the main text, it is utterly important   for systems with hard--wall boundary conditions that the thermodynamic limit $L \rightarrow \infty$ precedes the limit $\tau \rightarrow \infty$.
It means that the correct finite--size (FS) scaling should be carried out not for a single quantity but for the $\tau$--dependent Majorana autocorrelation function $\lambda(\tau)$. In order to efficiently perform such scaling, we first fit
$\lambda(\tau)$  for a given system length $L$ and then carry out the FS scaling for the fitting parameters.  It is convenient to start from a standard time--dependent correlation function  
\begin{eqnarray}
\langle \Gamma(t) \Gamma \rangle=\frac{1}{Z} \sum_{m,n}  \exp[i t(E_m-E_n)] |\langle n |\Gamma | m \rangle|^2,
\label{stdep}
\end{eqnarray}
It differs from the $\tau$--dependent autocorrelation functions $\langle \bar{\Gamma} \Gamma \rangle$ which utilizes time--averaging introduced in  Eq. (2) in the main text.  However, the relation between both functions can be  easily established
\begin{eqnarray}
\langle \bar{\Gamma} \Gamma \rangle&=&  \frac{1}{Z} \sum_{m,n}  \theta\left(\frac{1}{\tau} -|E_m-E_n| \right) |\langle n |\Gamma | m \rangle|^2 \nonumber  \\
&=& \frac{1}{Z} \sum_{m,n}  |\langle n |\Gamma | m \rangle|^2 \int_{-1/\tau}^{1/\tau} \mathrm{d} \omega \; \delta(\omega+E_m-E_n) \nonumber  \\
&=& \frac{1}{2 \pi} \int_{-1/\tau}^{1/\tau} \mathrm{d} \omega \int_{-\infty}^{\infty}  \mathrm{d} t e^{i \omega t} \langle \Gamma(t) \Gamma \rangle.
\label{srel}
\end{eqnarray}
One immediately finds the limit 
 \begin{eqnarray}
 \lim_{\tau \rightarrow \infty} \langle \bar{\Gamma} \Gamma \rangle= \frac{1}{Z} \sum_{m,n:E_m=E_n} |\langle n |\Gamma | m \rangle|^2 \nonumber. 
 \end{eqnarray}
This limit is exactly equal to the steady--state part of $\langle \Gamma(t) \Gamma \rangle$  [see Eq. (\ref{stdep})] which survives for arbitrarily large $t$. For finite $\tau$,  $\langle \bar{\Gamma} \Gamma \rangle$ represents 
the integrated low--frequency part of the Fourier transform of $\langle \Gamma(t) \Gamma \rangle$, as follows from the last line in Eq. (\ref{srel}). 
 We use   $\langle \bar{\Gamma} \Gamma \rangle$ instead of $\langle \Gamma(t) \Gamma \rangle$ because only the former function is monotonic. 
 It filters out the oscillations of  $\langle \Gamma(t) \Gamma \rangle$ but retains essential information about the asymptotic long--time behavior  \cite{annalen}.
 But most importantly, the efficiency of our method follows from that the specific time--averaging, $\bar{\Gamma}$, is an orthogonal projection also for $\tau < \infty$, i.e., $\langle \bar{\Gamma} \Gamma \rangle=
 \langle \bar{\Gamma}  \bar{\Gamma} \rangle $. The latter identity can be checked, by direct calculations. Namely, using  Eq. (3) from the main text, one obtains
\begin{eqnarray}
\langle \bar{\Gamma} \bar{\Gamma} \rangle&=&  \frac{1}{Z}\sum_{m,n} \left[  \theta\left(\frac{1}{\tau} -|E_m-E_n| \right)    \langle n |\Gamma | m \rangle  \right. \nonumber \\ 
&&\left . \theta\left(\frac{1}{\tau} -|E_n-E_m| \right)   \langle m |\Gamma | n \rangle \right].
\end{eqnarray}
Since both $\theta$--functions have equal arguments and $\theta^2(x)=\theta(x)$  one obtains a formula that is identical to the first line in Eq. (\ref{srel}).

\begin{figure}
\includegraphics[width=0.8 \columnwidth]{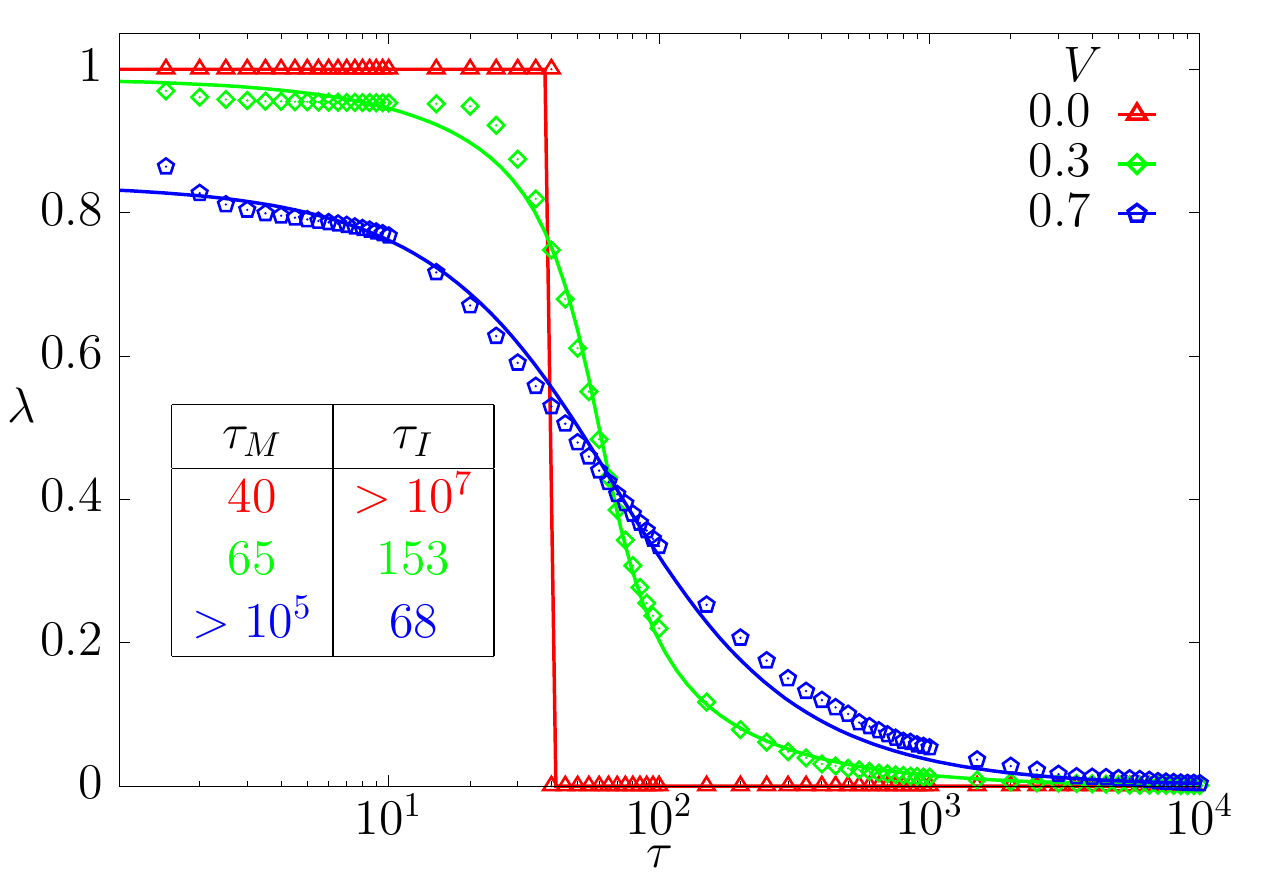}
\caption{
The Majorana autocorrelation function $\lambda$ for various  times $\tau$. Points show numerical results for  $L=12$, $\Delta=0.3$, $\mu=0$ and $W=V/2$ while continuous 
lines show best fits, given by Eq. (\ref{cfit}). The scattering times, $\tau_M$ and $\tau_I$, are shown in the table.}
\label{S5}
\end{figure} 
\begin{figure}
\centering
\includegraphics[width=\columnwidth]{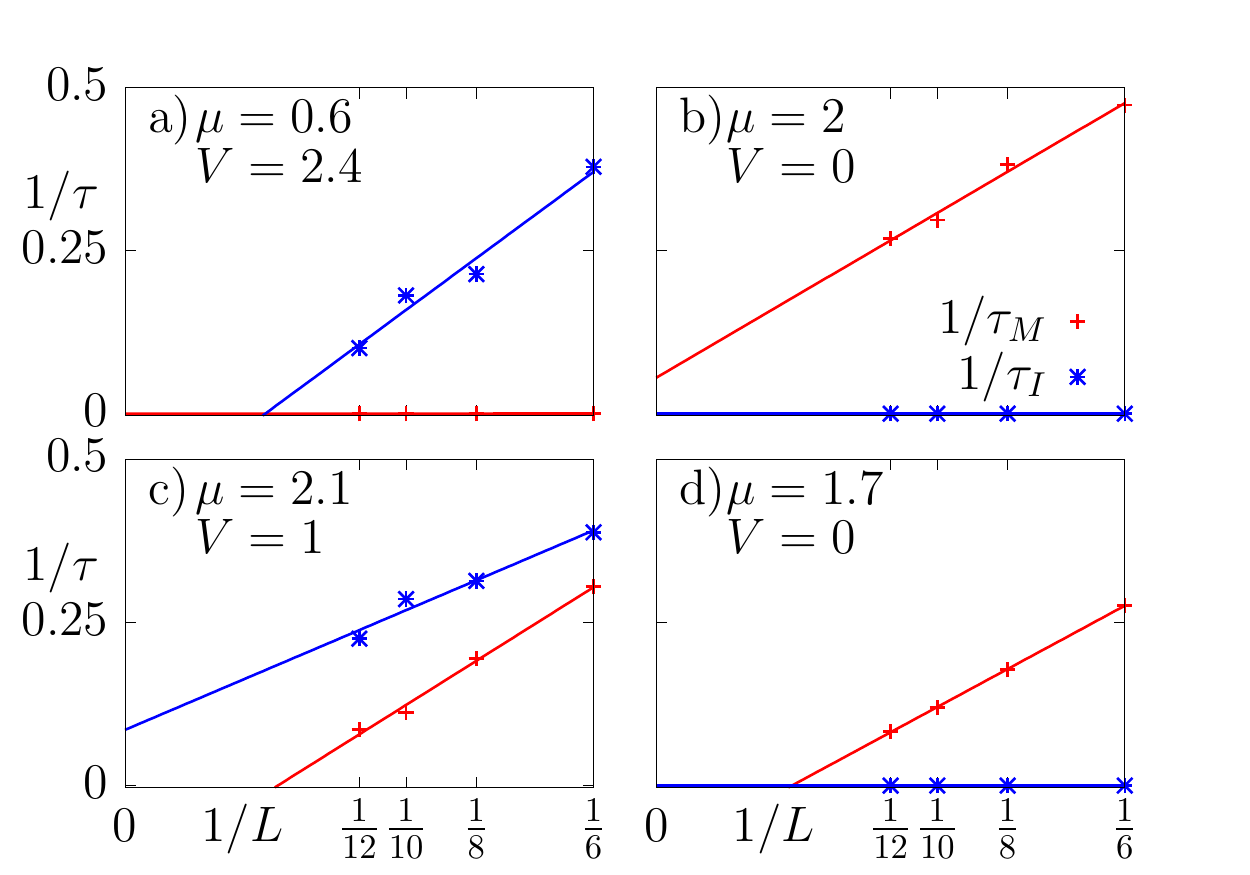}
\caption{
Representative examples for the finite--size scaling  of scattering times, $\tau_I$ and $\tau_M$ for $\Delta=1$, $W=0$.
}
\label{S6}
\end{figure} 

\begin{figure}
\centering
\includegraphics[width=\columnwidth]{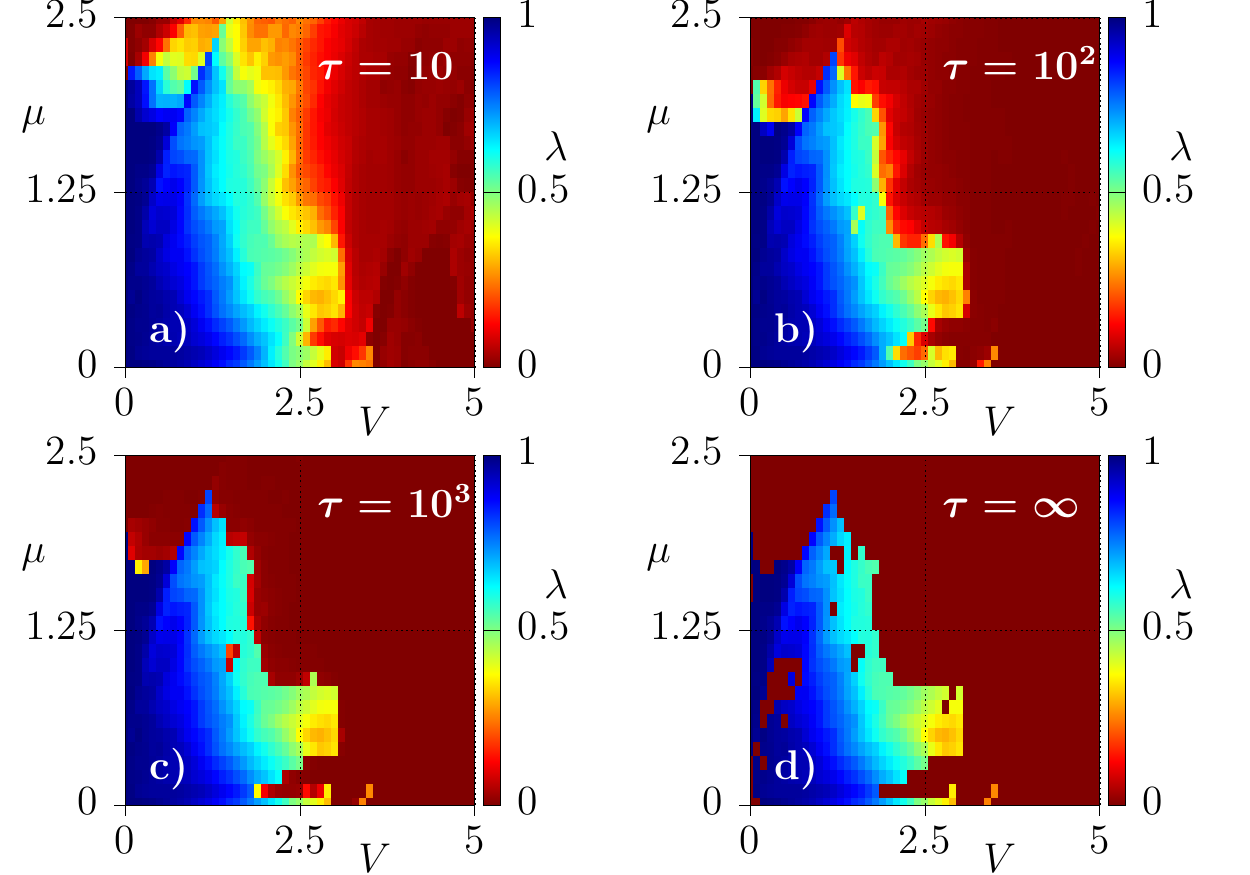}
\caption{
Extrapolated Majorana autocorrelation function, $\lim_{L \rightarrow \infty} \lambda$ for $\Delta=1$, $W=0$ various  times $\tau$.
 }
 \label{S7}
\end{figure} 

\begin{figure}
\centering
\includegraphics[width=\columnwidth]{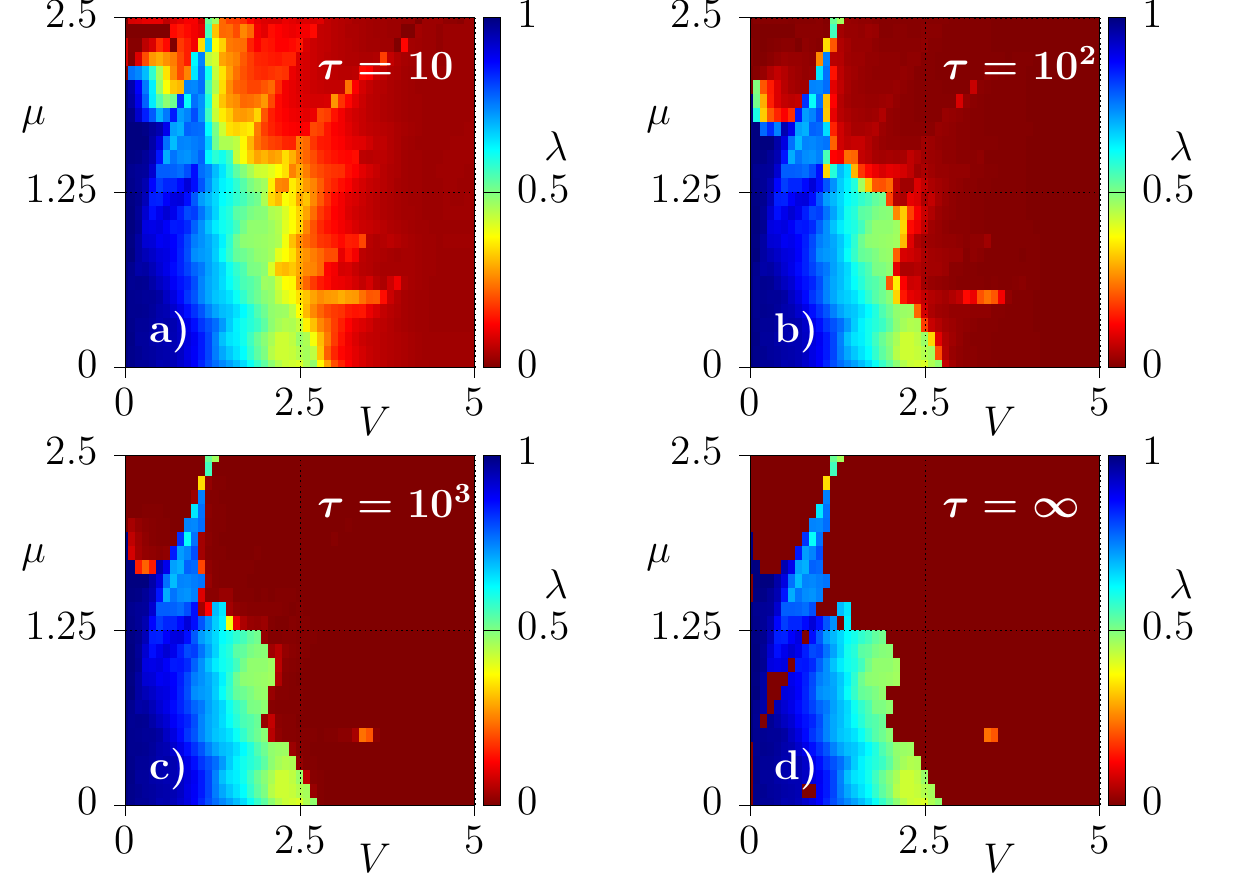}
\caption{ 
The same as in Fig. \ref{S7} but for $W=V/2$.
}
 \label{S8}
\end{figure}

In order to find the relevant fitting function for $\lambda(\tau)$, we notice that generic many--body interactions are expected to cause exponential decay of correlations functions. It holds true also
for perturbed integrable systems \cite{mm2}. For  $\langle \Gamma(t) \Gamma \rangle=\exp(-t/\tau_{I})$  one obtains from Eq. (\ref{srel}) 
 \begin{eqnarray}
\langle \bar{\Gamma} \Gamma \rangle&=&\frac{2}{\pi} \arctan \left( \frac{\tau_I}{\tau} \right). 
\label{stau}
\end{eqnarray}
However, such form is too simple to account for  {\it finite} noninteracting systems, where the Majorana lifetime, $\tau_M$, is limited by the overlap of Majorana modes  at two ends of the chain. 
Without interactions, the Majorana autocorrelation function  is a step function, as  shown in Fig. 1a in the main text. In order to accommodate this mechanism we have modified Eq. (\ref{stau})     
\begin{eqnarray}
\lambda_\mathrm{fit}(\tau)&=& \frac{C}{\pi} \left[ \arctan \left(\frac{\tau_{\mathrm{I}}}{\tau}-\frac{\tau_{\mathrm{I}}}{\tau_{M}} \right)  
+
 \arctan \left(\frac{\tau_{\mathrm{I}}}{\tau}+\frac{\tau_{\mathrm{I}}}{\tau_{M}} \right) \right].
\nonumber \\
\label{cfit}
\end{eqnarray}

The fitting function contains three parameters: $C$,  $\tau_\mathrm{I}$ and  $\tau_M$. Here, $1/\tau_M$ and $1/\tau_\mathrm{I}$ represent scattering rates due to the overlap of two Majorana modes and due to the many--body interactions, respectively. If the relaxation is dominated by the many--body interactions, then $\lim_{\tau_M \rightarrow \infty} \lambda_\mathrm{fit}(\tau)= \frac{2C}{\pi} \arctan \left( \frac{\tau_\mathrm{I}}{\tau} \right)$.
However, if the relaxation is due to overlap of two Majorana modes, then  $\lim_{\tau_\mathrm{I} \rightarrow \infty} \lambda_\mathrm{fit}(\tau)=  C \theta(\tau_M-\tau)$, in agreement with  Fig. 1a in the main text.
Figure  \ref{S5}  shows that  $\lambda(\tau)$ may be well fitted by Eq. (\ref{cfit}) also for intermediate cases, when both scattering mechanisms are important.   

In Fig. \ref{S6} we show a few representative examples for the finite size scaling of the scattering rates $1/\tau_\mathrm{I}$ and $1/\tau_M$.   One is mostly interested in the case when
both scattering rates vanish in the thermodynamic limit (see Figs. \ref{S6}a, \ref{S6}d) and  $\lim_{\tau \rightarrow \infty} \lim_{L \rightarrow \infty}   \lambda (\tau)= \lim_{L \rightarrow \infty} C$. 
Otherwise, $\lim_{\tau \rightarrow \infty} \lim_{L \rightarrow \infty}   \lambda (\tau)=0$  if one of the scattering times remain finite. In noninteracting system, the Majorana modes exist for 
$|\mu| < 2$ what is correctly reproduced by our approach as shown in Figs. \ref{S6}b and \ref{S6}d.

After the FS scaling has been accomplished,  one may study (approximate) results for  the Majorana autocorrelation function in the thermodynamic limit presented in Figs. \ref{S7} and \ref{S8}.  
These plots show $\lambda_\mathrm{fit}(\tau)$ where all fitting parameters are replaced by their extrapolated values.  
Such procedure unavoidably introduces errors. Consequently the irregular shape of the regime with Majorana modes most probably arises as a numerical artifacts. Nevertheless, it is rather evident 
that the information stored in the Majorana autocorrelation functions is at least partially retained for arbitrarily long times also for rather strong interactions $V \lesssim 2$.  
However, if the Majorana modes are strict integrals of motion then $\lim_{\tau \rightarrow \infty} \lim_{L \rightarrow \infty}   \lambda(\tau)= 1$.  Numerical results shown  in Figs. \ref{S7} and \ref{S8}  strongly suggest that in the presence of many--body interactions, the latter limit is always smaller than unity, even though it may be very close to this value. As argued in the main text, this implies  the presence of quasilocal strictly conserved operators $\lim_{\tau \rightarrow \infty} \bar{\Gamma}$, which
have large projection on strictly local Majorana modes, $\Gamma$.

The method has been demonstrated for a toy model of spinless fermions with $p$--wave pairing.  This model, however, is a prototype on which all current 1D realizations of Majorana physics are based \cite{PhysRevLett.105.077001,PhysRevLett.105.177002}. In real systems, the Zeeman splitting is used to break the Kramers degeneracy and create effectively  spinless fermions. The Rashba spin--orbit coupling combined with $s$--wave pairing induced by proximity to a conventional superconductor produces effective $p$--wave pairing. Then, in order to enter the topological phase that guarantees the presence of Majorana end modes, the parameters of the realistic model (Zeeman splitting $V_Z$, chemical potential $\mu$ and superconducting gap magnitude $\Delta$) have to satisfy exactly the same relation as derived for the Kitaev chain: $V_Z>\sqrt{\Delta^2+\mu^2}$ \cite{Franz2013}.
Moreover, the proposed method can  be straightforwardly applied to models of the semiconducting nanowires used in  real experiments. The only problem is that for spinful electrons the Hilbert space  is twice as large as in the Kitaev model, what would limit the maximum length of the system. And since the lifetime of MZM increases with the system size, we preferred our system to be as large as possible.

\bibliography{lib.bib}

\begin{thebibliography}{51}%
\makeatletter
\providecommand \@ifxundefined [1]{%
 \@ifx{#1\undefined}
}%
\providecommand \@ifnum [1]{%
 \ifnum #1\expandafter \@firstoftwo
 \else \expandafter \@secondoftwo
 \fi
}%
\providecommand \@ifx [1]{%
 \ifx #1\expandafter \@firstoftwo
 \else \expandafter \@secondoftwo
 \fi
}%
\providecommand \natexlab [1]{#1}%
\providecommand \enquote  [1]{``#1''}%
\providecommand \bibnamefont  [1]{#1}%
\providecommand \bibfnamefont [1]{#1}%
\providecommand \citenamefont [1]{#1}%
\providecommand \href@noop [0]{\@secondoftwo}%
\providecommand \href [0]{\begingroup \@sanitize@url \@href}%
\providecommand \@href[1]{\@@startlink{#1}\@@href}%
\providecommand \@@href[1]{\endgroup#1\@@endlink}%
\providecommand \@sanitize@url [0]{\catcode `\\12\catcode `\$12\catcode
  `\&12\catcode `\#12\catcode `\^12\catcode `\_12\catcode `\%12\relax}%
\providecommand \@@startlink[1]{}%
\providecommand \@@endlink[0]{}%
\providecommand \url  [0]{\begingroup\@sanitize@url \@url }%
\providecommand \@url [1]{\endgroup\@href {#1}{\urlprefix }}%
\providecommand \urlprefix  [0]{URL }%
\providecommand \Eprint [0]{\href }%
\providecommand \doibase [0]{http://dx.doi.org/}%
\providecommand \selectlanguage [0]{\@gobble}%
\providecommand \bibinfo  [0]{\@secondoftwo}%
\providecommand \bibfield  [0]{\@secondoftwo}%
\providecommand \translation [1]{[#1]}%
\providecommand \BibitemOpen [0]{}%
\providecommand \bibitemStop [0]{}%
\providecommand \bibitemNoStop [0]{.\EOS\space}%
\providecommand \EOS [0]{\spacefactor3000\relax}%
\providecommand \BibitemShut  [1]{\csname bibitem#1\endcsname}%
\let\auto@bib@innerbib\@empty
\bibitem [{\citenamefont {Aasen}\ \emph {et~al.}(2016)\citenamefont {Aasen},
  \citenamefont {Hell}, \citenamefont {Mishmash}, \citenamefont {Higginbotham},
  \citenamefont {Danon}, \citenamefont {Leijnse}, \citenamefont {Jespersen},
  \citenamefont {Folk}, \citenamefont {Marcus}, \citenamefont {Flensberg},\
  and\ \citenamefont {Alicea}}]{QC1}%
  \BibitemOpen
  \bibfield  {author} {\bibinfo {author} {\bibfnamefont {David}\ \bibnamefont
  {Aasen}}, \bibinfo {author} {\bibfnamefont {Michael}\ \bibnamefont {Hell}},
  \bibinfo {author} {\bibfnamefont {Ryan~V.}\ \bibnamefont {Mishmash}},
  \bibinfo {author} {\bibfnamefont {Andrew}\ \bibnamefont {Higginbotham}},
  \bibinfo {author} {\bibfnamefont {Jeroen}\ \bibnamefont {Danon}}, \bibinfo
  {author} {\bibfnamefont {Martin}\ \bibnamefont {Leijnse}}, \bibinfo {author}
  {\bibfnamefont {Thomas~S.}\ \bibnamefont {Jespersen}}, \bibinfo {author}
  {\bibfnamefont {Joshua~A.}\ \bibnamefont {Folk}}, \bibinfo {author}
  {\bibfnamefont {Charles~M.}\ \bibnamefont {Marcus}}, \bibinfo {author}
  {\bibfnamefont {Karsten}\ \bibnamefont {Flensberg}}, \ and\ \bibinfo {author}
  {\bibfnamefont {Jason}\ \bibnamefont {Alicea}},\ }\bibfield  {title}
  {\enquote {\bibinfo {title} {Milestones toward {M}ajorana-based quantum
  computing},}\ }\href {\doibase 10.1103/PhysRevX.6.031016} {\bibfield
  {journal} {\bibinfo  {journal} {Phys. Rev. X}\ }\textbf {\bibinfo {volume}
  {6}},\ \bibinfo {pages} {031016} (\bibinfo {year} {2016})}\BibitemShut
  {NoStop}%
\bibitem [{\citenamefont {Sarma}\ \emph {et~al.}(2015)\citenamefont {Sarma},
  \citenamefont {Freedman},\ and\ \citenamefont {Nayak}}]{QC2}%
  \BibitemOpen
  \bibfield  {author} {\bibinfo {author} {\bibfnamefont {Sankar~Das}\
  \bibnamefont {Sarma}}, \bibinfo {author} {\bibfnamefont {Michael}\
  \bibnamefont {Freedman}}, \ and\ \bibinfo {author} {\bibfnamefont {Chetan}\
  \bibnamefont {Nayak}},\ }\bibfield  {title} {\enquote {\bibinfo {title}
  {{M}ajorana zero modes and topological quantum computation},}\ }\href
  {\doibase 10.1038/npjqi.2015.1} {\bibfield  {journal} {\bibinfo  {journal}
  {npj Quantum Information}\ }\textbf {\bibinfo {volume} {1}},\ \bibinfo
  {pages} {15001} (\bibinfo {year} {2015})}\BibitemShut {NoStop}%
\bibitem [{\citenamefont {Karzig}\ \emph {et~al.}(2017)\citenamefont {Karzig},
  \citenamefont {Knapp}, \citenamefont {Lutchyn}, \citenamefont {Bonderson},
  \citenamefont {Hastings}, \citenamefont {Nayak}, \citenamefont {Alicea},
  \citenamefont {Flensberg}, \citenamefont {Plugge}, \citenamefont {Oreg},
  \citenamefont {Marcus},\ and\ \citenamefont {Freedman}}]{QC3}%
  \BibitemOpen
  \bibfield  {author} {\bibinfo {author} {\bibfnamefont {Torsten}\ \bibnamefont
  {Karzig}}, \bibinfo {author} {\bibfnamefont {Christina}\ \bibnamefont
  {Knapp}}, \bibinfo {author} {\bibfnamefont {Roman~M.}\ \bibnamefont
  {Lutchyn}}, \bibinfo {author} {\bibfnamefont {Parsa}\ \bibnamefont
  {Bonderson}}, \bibinfo {author} {\bibfnamefont {Matthew~B.}\ \bibnamefont
  {Hastings}}, \bibinfo {author} {\bibfnamefont {Chetan}\ \bibnamefont
  {Nayak}}, \bibinfo {author} {\bibfnamefont {Jason}\ \bibnamefont {Alicea}},
  \bibinfo {author} {\bibfnamefont {Karsten}\ \bibnamefont {Flensberg}},
  \bibinfo {author} {\bibfnamefont {Stephan}\ \bibnamefont {Plugge}}, \bibinfo
  {author} {\bibfnamefont {Yuval}\ \bibnamefont {Oreg}}, \bibinfo {author}
  {\bibfnamefont {Charles~M.}\ \bibnamefont {Marcus}}, \ and\ \bibinfo {author}
  {\bibfnamefont {Michael~H.}\ \bibnamefont {Freedman}},\ }\bibfield  {title}
  {\enquote {\bibinfo {title} {Scalable designs for
  quasiparticle-poisoning-protected topological quantum computation with
  {M}ajorana zero modes},}\ }\href {\doibase 10.1103/PhysRevB.95.235305}
  {\bibfield  {journal} {\bibinfo  {journal} {Phys. Rev. B}\ }\textbf {\bibinfo
  {volume} {95}},\ \bibinfo {pages} {235305} (\bibinfo {year}
  {2017})}\BibitemShut {NoStop}%
\bibitem [{\citenamefont {Plugge}\ \emph {et~al.}(2016)\citenamefont {Plugge},
  \citenamefont {Landau}, \citenamefont {Sela}, \citenamefont {Altland},
  \citenamefont {Flensberg},\ and\ \citenamefont {Egger}}]{QC4}%
  \BibitemOpen
  \bibfield  {author} {\bibinfo {author} {\bibfnamefont {S.}~\bibnamefont
  {Plugge}}, \bibinfo {author} {\bibfnamefont {L.~A.}\ \bibnamefont {Landau}},
  \bibinfo {author} {\bibfnamefont {E.}~\bibnamefont {Sela}}, \bibinfo {author}
  {\bibfnamefont {A.}~\bibnamefont {Altland}}, \bibinfo {author} {\bibfnamefont
  {K.}~\bibnamefont {Flensberg}}, \ and\ \bibinfo {author} {\bibfnamefont
  {R.}~\bibnamefont {Egger}},\ }\bibfield  {title} {\enquote {\bibinfo {title}
  {Roadmap to {M}ajorana surface codes},}\ }\href {\doibase
  10.1103/PhysRevB.94.174514} {\bibfield  {journal} {\bibinfo  {journal} {Phys.
  Rev. B}\ }\textbf {\bibinfo {volume} {94}},\ \bibinfo {pages} {174514}
  (\bibinfo {year} {2016})}\BibitemShut {NoStop}%
\bibitem [{\citenamefont {Akhmerov}(2010)}]{Akhmerov}%
  \BibitemOpen
  \bibfield  {author} {\bibinfo {author} {\bibfnamefont {A.~R.}\ \bibnamefont
  {Akhmerov}},\ }\bibfield  {title} {\enquote {\bibinfo {title} {Topological
  quantum computation away from the ground state using {M}ajorana fermions},}\
  }\href {\doibase 10.1103/PhysRevB.82.020509} {\bibfield  {journal} {\bibinfo
  {journal} {Phys. Rev. B}\ }\textbf {\bibinfo {volume} {82}},\ \bibinfo
  {pages} {020509} (\bibinfo {year} {2010})}\BibitemShut {NoStop}%
\bibitem [{\citenamefont {Lutchyn}\ \emph {et~al.}(2010)\citenamefont
  {Lutchyn}, \citenamefont {Sau},\ and\ \citenamefont
  {Das~Sarma}}]{PhysRevLett.105.077001}%
  \BibitemOpen
  \bibfield  {author} {\bibinfo {author} {\bibfnamefont {Roman~M.}\
  \bibnamefont {Lutchyn}}, \bibinfo {author} {\bibfnamefont {Jay~D.}\
  \bibnamefont {Sau}}, \ and\ \bibinfo {author} {\bibfnamefont
  {S.}~\bibnamefont {Das~Sarma}},\ }\bibfield  {title} {\enquote {\bibinfo
  {title} {Majorana fermions and a topological phase transition in
  semiconductor-superconductor heterostructures},}\ }\href {\doibase
  10.1103/PhysRevLett.105.077001} {\bibfield  {journal} {\bibinfo  {journal}
  {Phys. Rev. Lett.}\ }\textbf {\bibinfo {volume} {105}},\ \bibinfo {pages}
  {077001} (\bibinfo {year} {2010})}\BibitemShut {NoStop}%
\bibitem [{\citenamefont {Oreg}\ \emph {et~al.}(2010)\citenamefont {Oreg},
  \citenamefont {Refael},\ and\ \citenamefont {von
  Oppen}}]{PhysRevLett.105.177002}%
  \BibitemOpen
  \bibfield  {author} {\bibinfo {author} {\bibfnamefont {Yuval}\ \bibnamefont
  {Oreg}}, \bibinfo {author} {\bibfnamefont {Gil}\ \bibnamefont {Refael}}, \
  and\ \bibinfo {author} {\bibfnamefont {Felix}\ \bibnamefont {von Oppen}},\
  }\bibfield  {title} {\enquote {\bibinfo {title} {Helical liquids and majorana
  bound states in quantum wires},}\ }\href {\doibase
  10.1103/PhysRevLett.105.177002} {\bibfield  {journal} {\bibinfo  {journal}
  {Phys. Rev. Lett.}\ }\textbf {\bibinfo {volume} {105}},\ \bibinfo {pages}
  {177002} (\bibinfo {year} {2010})}\BibitemShut {NoStop}%
\bibitem [{\citenamefont {Franz}(2013)}]{Franz2013}%
  \BibitemOpen
  \bibfield  {author} {\bibinfo {author} {\bibfnamefont {Marcel}\ \bibnamefont
  {Franz}},\ }\bibfield  {title} {\enquote {\bibinfo {title} {Majorana's
  wires},}\ }\href {\doibase 10.1038/nnano.2013.33} {\bibfield  {journal}
  {\bibinfo  {journal} {Nat. Nanotechnol}\ }\textbf {\bibinfo {volume} {8}},\
  \bibinfo {pages} {149--152} (\bibinfo {year} {2013})}\BibitemShut {NoStop}%
\bibitem [{\citenamefont {Mourik}\ \emph {et~al.}(2012)\citenamefont {Mourik},
  \citenamefont {Zuo}, \citenamefont {Frolov}, \citenamefont {Plissard},
  \citenamefont {Bakkers},\ and\ \citenamefont {Kouwenhoven}}]{Mourik1003}%
  \BibitemOpen
  \bibfield  {author} {\bibinfo {author} {\bibfnamefont {V.}~\bibnamefont
  {Mourik}}, \bibinfo {author} {\bibfnamefont {K.}~\bibnamefont {Zuo}},
  \bibinfo {author} {\bibfnamefont {S.~M.}\ \bibnamefont {Frolov}}, \bibinfo
  {author} {\bibfnamefont {S.~R.}\ \bibnamefont {Plissard}}, \bibinfo {author}
  {\bibfnamefont {E.~P. A.~M.}\ \bibnamefont {Bakkers}}, \ and\ \bibinfo
  {author} {\bibfnamefont {L.~P.}\ \bibnamefont {Kouwenhoven}},\ }\bibfield
  {title} {\enquote {\bibinfo {title} {Signatures of {M}ajorana fermions in
  hybrid superconductor-semiconductor nanowire devices},}\ }\href {\doibase
  10.1126/science.1222360} {\bibfield  {journal} {\bibinfo  {journal}
  {Science}\ }\textbf {\bibinfo {volume} {336}},\ \bibinfo {pages} {1003}
  (\bibinfo {year} {2012})}\BibitemShut {NoStop}%
\bibitem [{\citenamefont {Nadj-Perge}\ \emph {et~al.}(2014)\citenamefont
  {Nadj-Perge}, \citenamefont {Drozdov}, \citenamefont {Li}, \citenamefont
  {Chen}, \citenamefont {Jeon}, \citenamefont {Seo}, \citenamefont {MacDonald},
  \citenamefont {Bernevig},\ and\ \citenamefont {Yazdani}}]{Nadj-Perge602}%
  \BibitemOpen
  \bibfield  {author} {\bibinfo {author} {\bibfnamefont {Stevan}\ \bibnamefont
  {Nadj-Perge}}, \bibinfo {author} {\bibfnamefont {Ilya~K.}\ \bibnamefont
  {Drozdov}}, \bibinfo {author} {\bibfnamefont {Jian}\ \bibnamefont {Li}},
  \bibinfo {author} {\bibfnamefont {Hua}\ \bibnamefont {Chen}}, \bibinfo
  {author} {\bibfnamefont {Sangjun}\ \bibnamefont {Jeon}}, \bibinfo {author}
  {\bibfnamefont {Jungpil}\ \bibnamefont {Seo}}, \bibinfo {author}
  {\bibfnamefont {Allan~H.}\ \bibnamefont {MacDonald}}, \bibinfo {author}
  {\bibfnamefont {B.~Andrei}\ \bibnamefont {Bernevig}}, \ and\ \bibinfo
  {author} {\bibfnamefont {Ali}\ \bibnamefont {Yazdani}},\ }\bibfield  {title}
  {\enquote {\bibinfo {title} {Observation of {M}ajorana fermions in
  ferromagnetic atomic chains on a superconductor},}\ }\href {\doibase
  10.1126/science.1259327} {\bibfield  {journal} {\bibinfo  {journal}
  {Science}\ }\textbf {\bibinfo {volume} {346}},\ \bibinfo {pages} {602}
  (\bibinfo {year} {2014})}\BibitemShut {NoStop}%
\bibitem [{\citenamefont {Pawlak}\ \emph {et~al.}(2016)\citenamefont {Pawlak},
  \citenamefont {Kisiel}, \citenamefont {Klinovaja}, \citenamefont {Meier},
  \citenamefont {Kawai}, \citenamefont {Glatzel}, \citenamefont {Loss},\ and\
  \citenamefont {Meyer}}]{Pawlak2016}%
  \BibitemOpen
  \bibfield  {author} {\bibinfo {author} {\bibfnamefont {R{\'e}my}\
  \bibnamefont {Pawlak}}, \bibinfo {author} {\bibfnamefont {Marcin}\
  \bibnamefont {Kisiel}}, \bibinfo {author} {\bibfnamefont {Jelena}\
  \bibnamefont {Klinovaja}}, \bibinfo {author} {\bibfnamefont {Tobias}\
  \bibnamefont {Meier}}, \bibinfo {author} {\bibfnamefont {Shigeki}\
  \bibnamefont {Kawai}}, \bibinfo {author} {\bibfnamefont {Thilo}\ \bibnamefont
  {Glatzel}}, \bibinfo {author} {\bibfnamefont {Daniel}\ \bibnamefont {Loss}},
  \ and\ \bibinfo {author} {\bibfnamefont {Ernst}\ \bibnamefont {Meyer}},\
  }\bibfield  {title} {\enquote {\bibinfo {title} {Probing atomic structure and
  {M}ajorana wavefunctions in mono-atomic fe chains on superconducting pb
  surface},}\ }\href {http://dx.doi.org/10.1038/npjqi.2016.35} {\bibfield
  {journal} {\bibinfo  {journal} {npj Quantum Information}\ }\textbf {\bibinfo
  {volume} {2}},\ \bibinfo {pages} {16035} (\bibinfo {year}
  {2016})}\BibitemShut {NoStop}%
\bibitem [{\citenamefont {Ruby}\ \emph {et~al.}(2015)\citenamefont {Ruby},
  \citenamefont {Pientka}, \citenamefont {Peng}, \citenamefont {von Oppen},
  \citenamefont {Heinrich},\ and\ \citenamefont {Franke}}]{Ruby2015}%
  \BibitemOpen
  \bibfield  {author} {\bibinfo {author} {\bibfnamefont {Michael}\ \bibnamefont
  {Ruby}}, \bibinfo {author} {\bibfnamefont {Falko}\ \bibnamefont {Pientka}},
  \bibinfo {author} {\bibfnamefont {Yang}\ \bibnamefont {Peng}}, \bibinfo
  {author} {\bibfnamefont {Felix}\ \bibnamefont {von Oppen}}, \bibinfo {author}
  {\bibfnamefont {Benjamin~W.}\ \bibnamefont {Heinrich}}, \ and\ \bibinfo
  {author} {\bibfnamefont {Katharina~J.}\ \bibnamefont {Franke}},\ }\bibfield
  {title} {\enquote {\bibinfo {title} {End states and subgap structure in
  proximity-coupled chains of magnetic adatoms},}\ }\href {\doibase
  10.1103/PhysRevLett.115.197204} {\bibfield  {journal} {\bibinfo  {journal}
  {Phys. Rev. Lett.}\ }\textbf {\bibinfo {volume} {115}},\ \bibinfo {pages}
  {197204} (\bibinfo {year} {2015})}\BibitemShut {NoStop}%
\bibitem [{\citenamefont {Haldane}(1981)}]{Haldane}%
  \BibitemOpen
  \bibfield  {author} {\bibinfo {author} {\bibfnamefont {D.~M.}\ \bibnamefont
  {Haldane}},\ }\bibfield  {title} {\enquote {\bibinfo {title} {'{L}uttinger
  liquid theory' of one-dimensional quantum fluids. {I}. {P}roperties of the
  {L}uttinger model and their extension to the general 1{D} interacting
  spinless {F}ermi gas},}\ }\href
  {http://stacks.iop.org/0022-3719/14/i=19/a=010} {\bibfield  {journal}
  {\bibinfo  {journal} {J. Physics C.}\ }\textbf {\bibinfo {volume} {14}},\
  \bibinfo {pages} {2585} (\bibinfo {year} {1981})}\BibitemShut {NoStop}%
\bibitem [{\citenamefont {Gangadharaiah}\ \emph {et~al.}(2011)\citenamefont
  {Gangadharaiah}, \citenamefont {Braunecker}, \citenamefont {Simon},\ and\
  \citenamefont {Loss}}]{Gangadharaiah}%
  \BibitemOpen
  \bibfield  {author} {\bibinfo {author} {\bibfnamefont {Suhas}\ \bibnamefont
  {Gangadharaiah}}, \bibinfo {author} {\bibfnamefont {Bernd}\ \bibnamefont
  {Braunecker}}, \bibinfo {author} {\bibfnamefont {Pascal}\ \bibnamefont
  {Simon}}, \ and\ \bibinfo {author} {\bibfnamefont {Daniel}\ \bibnamefont
  {Loss}},\ }\bibfield  {title} {\enquote {\bibinfo {title} {{M}ajorana edge
  states in interacting one-dimensional systems},}\ }\href {\doibase
  10.1103/PhysRevLett.107.036801} {\bibfield  {journal} {\bibinfo  {journal}
  {Phys. Rev. Lett.}\ }\textbf {\bibinfo {volume} {107}},\ \bibinfo {pages}
  {036801} (\bibinfo {year} {2011})}\BibitemShut {NoStop}%
\bibitem [{\citenamefont {Manolescu}\ \emph {et~al.}(2014)\citenamefont
  {Manolescu}, \citenamefont {Marinescu},\ and\ \citenamefont
  {Stanescu}}]{superconductor_driven}%
  \BibitemOpen
  \bibfield  {author} {\bibinfo {author} {\bibfnamefont {A.}~\bibnamefont
  {Manolescu}}, \bibinfo {author} {\bibfnamefont {D.~C.}\ \bibnamefont
  {Marinescu}}, \ and\ \bibinfo {author} {\bibfnamefont {T.~D.}\ \bibnamefont
  {Stanescu}},\ }\bibfield  {title} {\enquote {\bibinfo {title} {Coulomb
  interaction effects on the {M}ajorana states in quantum wires},}\ }\href
  {http://stacks.iop.org/0953-8984/26/i=17/a=172203} {\bibfield  {journal}
  {\bibinfo  {journal} {J. Physics. Condens. Matter}\ }\textbf {\bibinfo
  {volume} {26}},\ \bibinfo {pages} {172203} (\bibinfo {year}
  {2014})}\BibitemShut {NoStop}%
\bibitem [{\citenamefont {Vuik}\ \emph {et~al.}(2016)\citenamefont {Vuik},
  \citenamefont {Eeltink}, \citenamefont {Akhmerov},\ and\ \citenamefont
  {Wimmer}}]{envir}%
  \BibitemOpen
  \bibfield  {author} {\bibinfo {author} {\bibfnamefont {A.}~\bibnamefont
  {Vuik}}, \bibinfo {author} {\bibfnamefont {D.}~\bibnamefont {Eeltink}},
  \bibinfo {author} {\bibfnamefont {A.~R.}\ \bibnamefont {Akhmerov}}, \ and\
  \bibinfo {author} {\bibfnamefont {M}~\bibnamefont {Wimmer}},\ }\bibfield
  {title} {\enquote {\bibinfo {title} {Effects of the electrostatic environment
  on the {M}ajorana nanowire devices},}\ }\href
  {http://stacks.iop.org/1367-2630/18/i=3/a=033013} {\bibfield  {journal}
  {\bibinfo  {journal} {New J. Phys.}\ }\textbf {\bibinfo {volume} {18}},\
  \bibinfo {pages} {033013} (\bibinfo {year} {2016})}\BibitemShut {NoStop}%
\bibitem [{\citenamefont {Dom{\'\i}nguez}\ \emph
  {et~al.}(2017{\natexlab{a}})\citenamefont {Dom{\'\i}nguez}, \citenamefont
  {Cayao}, \citenamefont {San-Jose}, \citenamefont {Aguado}, \citenamefont
  {Yeyati},\ and\ \citenamefont {Prada}}]{intdod}%
  \BibitemOpen
  \bibfield  {author} {\bibinfo {author} {\bibfnamefont {Fernando}\
  \bibnamefont {Dom{\'\i}nguez}}, \bibinfo {author} {\bibfnamefont {Jorge}\
  \bibnamefont {Cayao}}, \bibinfo {author} {\bibfnamefont {Pablo}\ \bibnamefont
  {San-Jose}}, \bibinfo {author} {\bibfnamefont {Ram{\'o}n}\ \bibnamefont
  {Aguado}}, \bibinfo {author} {\bibfnamefont {Alfredo~Levy}\ \bibnamefont
  {Yeyati}}, \ and\ \bibinfo {author} {\bibfnamefont {Elsa}\ \bibnamefont
  {Prada}},\ }\bibfield  {title} {\enquote {\bibinfo {title} {Zero-energy
  pinning from interactions in majorana nanowires},}\ }\href {\doibase
  10.1038/s41535-017-0012-0} {\bibfield  {journal} {\bibinfo  {journal} {npj
  Quantum Materials}\ }\textbf {\bibinfo {volume} {2}},\ \bibinfo {pages} {13}
  (\bibinfo {year} {2017}{\natexlab{a}})}\BibitemShut {NoStop}%
\bibitem [{\citenamefont {Ma\ifmmode~\acute{s}\else \'{s}\fi{}ka}\ \emph
  {et~al.}(2017)\citenamefont {Ma\ifmmode~\acute{s}\else \'{s}\fi{}ka},
  \citenamefont {Gorczyca-Goraj}, \citenamefont {Tworzyd\l{}o},\ and\
  \citenamefont {Doma\ifmmode~\acute{n}\else \'{n}\fi{}ski}}]{Maska}%
  \BibitemOpen
  \bibfield  {author} {\bibinfo {author} {\bibfnamefont {Maciej~M.}\
  \bibnamefont {Ma\ifmmode~\acute{s}\else \'{s}\fi{}ka}}, \bibinfo {author}
  {\bibfnamefont {Anna}\ \bibnamefont {Gorczyca-Goraj}}, \bibinfo {author}
  {\bibfnamefont {Jakub}\ \bibnamefont {Tworzyd\l{}o}}, \ and\ \bibinfo
  {author} {\bibfnamefont {Tadeusz}\ \bibnamefont {Doma\ifmmode~\acute{n}\else
  \'{n}\fi{}ski}},\ }\bibfield  {title} {\enquote {\bibinfo {title} {{M}ajorana
  quasiparticles of an inhomogeneous rashba chain},}\ }\href {\doibase
  10.1103/PhysRevB.95.045429} {\bibfield  {journal} {\bibinfo  {journal} {Phys.
  Rev. B}\ }\textbf {\bibinfo {volume} {95}},\ \bibinfo {pages} {045429}
  (\bibinfo {year} {2017})}\BibitemShut {NoStop}%
\bibitem [{\citenamefont {Lutchyn}\ \emph {et~al.}(2011)\citenamefont
  {Lutchyn}, \citenamefont {Stanescu},\ and\ \citenamefont
  {Das~Sarma}}]{Lutchyn2011}%
  \BibitemOpen
  \bibfield  {author} {\bibinfo {author} {\bibfnamefont {Roman~M.}\
  \bibnamefont {Lutchyn}}, \bibinfo {author} {\bibfnamefont {Tudor~D.}\
  \bibnamefont {Stanescu}}, \ and\ \bibinfo {author} {\bibfnamefont
  {S.}~\bibnamefont {Das~Sarma}},\ }\bibfield  {title} {\enquote {\bibinfo
  {title} {Search for {M}ajorana fermions in multiband semiconducting
  nanowires},}\ }\href {\doibase 10.1103/PhysRevLett.106.127001} {\bibfield
  {journal} {\bibinfo  {journal} {Phys. Rev. Lett.}\ }\textbf {\bibinfo
  {volume} {106}},\ \bibinfo {pages} {127001} (\bibinfo {year}
  {2011})}\BibitemShut {NoStop}%
\bibitem [{\citenamefont {Akhmerov}\ \emph {et~al.}(2011)\citenamefont
  {Akhmerov}, \citenamefont {Dahlhaus}, \citenamefont {Hassler}, \citenamefont
  {Wimmer},\ and\ \citenamefont {Beenakker}}]{Akhmerov2011}%
  \BibitemOpen
  \bibfield  {author} {\bibinfo {author} {\bibfnamefont {A.~R.}\ \bibnamefont
  {Akhmerov}}, \bibinfo {author} {\bibfnamefont {J.~P.}\ \bibnamefont
  {Dahlhaus}}, \bibinfo {author} {\bibfnamefont {F.}~\bibnamefont {Hassler}},
  \bibinfo {author} {\bibfnamefont {M.}~\bibnamefont {Wimmer}}, \ and\ \bibinfo
  {author} {\bibfnamefont {C.~W.~J.}\ \bibnamefont {Beenakker}},\ }\bibfield
  {title} {\enquote {\bibinfo {title} {Quantized conductance at the {M}ajorana
  phase transition in a disordered superconducting wire},}\ }\href {\doibase
  10.1103/PhysRevLett.106.057001} {\bibfield  {journal} {\bibinfo  {journal}
  {Phys. Rev. Lett.}\ }\textbf {\bibinfo {volume} {106}},\ \bibinfo {pages}
  {057001} (\bibinfo {year} {2011})}\BibitemShut {NoStop}%
\bibitem [{\citenamefont {Dom{\'\i}nguez}\ \emph
  {et~al.}(2017{\natexlab{b}})\citenamefont {Dom{\'\i}nguez}, \citenamefont
  {Cayao}, \citenamefont {San-Jose}, \citenamefont {Aguado}, \citenamefont
  {Yeyati},\ and\ \citenamefont {Prada}}]{pinning}%
  \BibitemOpen
  \bibfield  {author} {\bibinfo {author} {\bibfnamefont {Fernando}\
  \bibnamefont {Dom{\'\i}nguez}}, \bibinfo {author} {\bibfnamefont {Jorge}\
  \bibnamefont {Cayao}}, \bibinfo {author} {\bibfnamefont {Pablo}\ \bibnamefont
  {San-Jose}}, \bibinfo {author} {\bibfnamefont {Ram{\'o}n}\ \bibnamefont
  {Aguado}}, \bibinfo {author} {\bibfnamefont {Alfredo~Levy}\ \bibnamefont
  {Yeyati}}, \ and\ \bibinfo {author} {\bibfnamefont {Elsa}\ \bibnamefont
  {Prada}},\ }\bibfield  {title} {\enquote {\bibinfo {title} {Zero-energy
  pinning from interactions in {M}ajorana nanowires},}\ }\href {\doibase
  10.1038/s41535-017-0012-0} {\bibfield  {journal} {\bibinfo  {journal} {npj
  Quantum Materials}\ }\textbf {\bibinfo {volume} {2}},\ \bibinfo {pages} {13}
  (\bibinfo {year} {2017}{\natexlab{b}})}\BibitemShut {NoStop}%
\bibitem [{\citenamefont {Stoudenmire}\ \emph {et~al.}(2011)\citenamefont
  {Stoudenmire}, \citenamefont {Alicea}, \citenamefont {Starykh},\ and\
  \citenamefont {Fisher}}]{Stoudenmire}%
  \BibitemOpen
  \bibfield  {author} {\bibinfo {author} {\bibfnamefont {E.~M.}\ \bibnamefont
  {Stoudenmire}}, \bibinfo {author} {\bibfnamefont {Jason}\ \bibnamefont
  {Alicea}}, \bibinfo {author} {\bibfnamefont {Oleg~A.}\ \bibnamefont
  {Starykh}}, \ and\ \bibinfo {author} {\bibfnamefont {Matthew~P.A.}\
  \bibnamefont {Fisher}},\ }\bibfield  {title} {\enquote {\bibinfo {title}
  {Interaction effects in topological superconducting wires supporting
  {M}ajorana fermions},}\ }\href {\doibase 10.1103/PhysRevB.84.014503}
  {\bibfield  {journal} {\bibinfo  {journal} {Phys. Rev. B}\ }\textbf {\bibinfo
  {volume} {84}},\ \bibinfo {pages} {014503} (\bibinfo {year}
  {2011})}\BibitemShut {NoStop}%
\bibitem [{\citenamefont {Gergs}\ \emph {et~al.}(2016)\citenamefont {Gergs},
  \citenamefont {Fritz},\ and\ \citenamefont {Schuricht}}]{Gergs}%
  \BibitemOpen
  \bibfield  {author} {\bibinfo {author} {\bibfnamefont {Niklas~M.}\
  \bibnamefont {Gergs}}, \bibinfo {author} {\bibfnamefont {Lars}\ \bibnamefont
  {Fritz}}, \ and\ \bibinfo {author} {\bibfnamefont {Dirk}\ \bibnamefont
  {Schuricht}},\ }\bibfield  {title} {\enquote {\bibinfo {title} {Topological
  order in the {K}itaev/{M}ajorana chain in the presence of disorder and
  interactions},}\ }\href {\doibase 10.1103/PhysRevB.93.075129} {\bibfield
  {journal} {\bibinfo  {journal} {Phys. Rev. B}\ }\textbf {\bibinfo {volume}
  {93}},\ \bibinfo {pages} {075129} (\bibinfo {year} {2016})}\BibitemShut
  {NoStop}%
\bibitem [{\citenamefont {Hassler}\ and\ \citenamefont
  {Schuricht}(2012)}]{Hassler2012}%
  \BibitemOpen
  \bibfield  {author} {\bibinfo {author} {\bibfnamefont {Fabian}\ \bibnamefont
  {Hassler}}\ and\ \bibinfo {author} {\bibfnamefont {Dirk}\ \bibnamefont
  {Schuricht}},\ }\bibfield  {title} {\enquote {\bibinfo {title} {Strongly
  interacting {M}ajorana modes in an array of josephson junctions},}\ }\href
  {http://stacks.iop.org/1367-2630/14/i=12/a=125018} {\bibfield  {journal}
  {\bibinfo  {journal} {New J. Phys.}\ }\textbf {\bibinfo {volume} {14}},\
  \bibinfo {pages} {125018} (\bibinfo {year} {2012})}\BibitemShut {NoStop}%
\bibitem [{\citenamefont {Nayak}\ \emph {et~al.}(2008)\citenamefont {Nayak},
  \citenamefont {Simon}, \citenamefont {Stern}, \citenamefont {Freedman},\ and\
  \citenamefont {Das~Sarma}}]{RevModPhys.80.1083}%
  \BibitemOpen
  \bibfield  {author} {\bibinfo {author} {\bibfnamefont {Chetan}\ \bibnamefont
  {Nayak}}, \bibinfo {author} {\bibfnamefont {Steven~H.}\ \bibnamefont
  {Simon}}, \bibinfo {author} {\bibfnamefont {Ady}\ \bibnamefont {Stern}},
  \bibinfo {author} {\bibfnamefont {Michael}\ \bibnamefont {Freedman}}, \ and\
  \bibinfo {author} {\bibfnamefont {Sankar}\ \bibnamefont {Das~Sarma}},\
  }\bibfield  {title} {\enquote {\bibinfo {title} {Non-abelian anyons and
  topological quantum computation},}\ }\href {\doibase
  10.1103/RevModPhys.80.1083} {\bibfield  {journal} {\bibinfo  {journal} {Rev.
  Mod. Phys.}\ }\textbf {\bibinfo {volume} {80}},\ \bibinfo {pages} {1083}
  (\bibinfo {year} {2008})}\BibitemShut {NoStop}%
\bibitem [{\citenamefont {Clarke}\ \emph {et~al.}(2011)\citenamefont {Clarke},
  \citenamefont {Sau},\ and\ \citenamefont {Tewari}}]{nonabelian}%
  \BibitemOpen
  \bibfield  {author} {\bibinfo {author} {\bibfnamefont {David~J.}\
  \bibnamefont {Clarke}}, \bibinfo {author} {\bibfnamefont {Jay~D.}\
  \bibnamefont {Sau}}, \ and\ \bibinfo {author} {\bibfnamefont {Sumanta}\
  \bibnamefont {Tewari}},\ }\bibfield  {title} {\enquote {\bibinfo {title}
  {Majorana fermion exchange in quasi-one-dimensional networks},}\ }\href
  {\doibase 10.1103/PhysRevB.84.035120} {\bibfield  {journal} {\bibinfo
  {journal} {Phys. Rev. B}\ }\textbf {\bibinfo {volume} {84}},\ \bibinfo
  {pages} {035120} (\bibinfo {year} {2011})}\BibitemShut {NoStop}%
\bibitem [{\citenamefont {Kitaev}(2001)}]{kitaev}%
  \BibitemOpen
  \bibfield  {author} {\bibinfo {author} {\bibfnamefont {A~Yu}\ \bibnamefont
  {Kitaev}},\ }\bibfield  {title} {\enquote {\bibinfo {title} {Unpaired
  {M}ajorana fermions in quantum wires},}\ }\href
  {http://stacks.iop.org/1063-7869/44/i=10S/a=S29} {\bibfield  {journal}
  {\bibinfo  {journal} {Phys. Usp.}\ }\textbf {\bibinfo {volume} {44}},\
  \bibinfo {pages} {131} (\bibinfo {year} {2001})}\BibitemShut {NoStop}%
\bibitem [{\citenamefont {Kells}(2015)}]{Kells}%
  \BibitemOpen
  \bibfield  {author} {\bibinfo {author} {\bibfnamefont {G.}~\bibnamefont
  {Kells}},\ }\bibfield  {title} {\enquote {\bibinfo {title} {Many-body
  {M}ajorana operators and the equivalence of parity sectors},}\ }\href
  {\doibase 10.1103/PhysRevB.92.081401} {\bibfield  {journal} {\bibinfo
  {journal} {Phys. Rev. B}\ }\textbf {\bibinfo {volume} {92}},\ \bibinfo
  {pages} {081401} (\bibinfo {year} {2015})}\BibitemShut {NoStop}%
\bibitem [{\citenamefont {Ng}(2015)}]{Ng}%
  \BibitemOpen
  \bibfield  {author} {\bibinfo {author} {\bibfnamefont {H.~T.}\ \bibnamefont
  {Ng}},\ }\bibfield  {title} {\enquote {\bibinfo {title} {Decoherence of
  interacting {M}ajorana modes},}\ }\href {http://dx.doi.org/10.1038/srep12530}
  {\bibfield  {journal} {\bibinfo  {journal} {Sci. Rep.}\ }\textbf {\bibinfo
  {volume} {5}},\ \bibinfo {pages} {12530} (\bibinfo {year}
  {2015})}\BibitemShut {NoStop}%
\bibitem [{\citenamefont {Su}\ \emph {et~al.}(2016)\citenamefont {Su},
  \citenamefont {Chen}, \citenamefont {Shao}, \citenamefont {Sheng},\ and\
  \citenamefont {Xing}}]{Su}%
  \BibitemOpen
  \bibfield  {author} {\bibinfo {author} {\bibfnamefont {W.}~\bibnamefont
  {Su}}, \bibinfo {author} {\bibfnamefont {M.~N.}\ \bibnamefont {Chen}},
  \bibinfo {author} {\bibfnamefont {L.~B.}\ \bibnamefont {Shao}}, \bibinfo
  {author} {\bibfnamefont {L.}~\bibnamefont {Sheng}}, \ and\ \bibinfo {author}
  {\bibfnamefont {D.~Y.}\ \bibnamefont {Xing}},\ }\bibfield  {title} {\enquote
  {\bibinfo {title} {Electron-electron interaction effects in floquet
  topological superconducting chains: Suppression of topological edge states
  and crossover from weak to strong chaos},}\ }\href {\doibase
  10.1103/PhysRevB.94.075145} {\bibfield  {journal} {\bibinfo  {journal} {Phys.
  Rev. B}\ }\textbf {\bibinfo {volume} {94}},\ \bibinfo {pages} {075145}
  (\bibinfo {year} {2016})}\BibitemShut {NoStop}%
\bibitem [{\citenamefont {Chan}\ \emph {et~al.}(2015)\citenamefont {Chan},
  \citenamefont {Chiu},\ and\ \citenamefont {Sun}}]{Chan}%
  \BibitemOpen
  \bibfield  {author} {\bibinfo {author} {\bibfnamefont {Y.-H.}\ \bibnamefont
  {Chan}}, \bibinfo {author} {\bibfnamefont {Ching-Kai}\ \bibnamefont {Chiu}},
  \ and\ \bibinfo {author} {\bibfnamefont {Kuei}\ \bibnamefont {Sun}},\
  }\bibfield  {title} {\enquote {\bibinfo {title} {Multiple signatures of
  topological transitions for interacting fermions in chain lattices},}\ }\href
  {\doibase 10.1103/PhysRevB.92.104514} {\bibfield  {journal} {\bibinfo
  {journal} {Phys. Rev. B}\ }\textbf {\bibinfo {volume} {92}},\ \bibinfo
  {pages} {104514} (\bibinfo {year} {2015})}\BibitemShut {NoStop}%
\bibitem [{\citenamefont {Hofmann}\ \emph {et~al.}(2016)\citenamefont
  {Hofmann}, \citenamefont {Assaad},\ and\ \citenamefont {Schnyder}}]{Hofmann}%
  \BibitemOpen
  \bibfield  {author} {\bibinfo {author} {\bibfnamefont {Johannes~S.}\
  \bibnamefont {Hofmann}}, \bibinfo {author} {\bibfnamefont {Fakher~F.}\
  \bibnamefont {Assaad}}, \ and\ \bibinfo {author} {\bibfnamefont {Andreas~P.}\
  \bibnamefont {Schnyder}},\ }\bibfield  {title} {\enquote {\bibinfo {title}
  {Edge instabilities of topological superconductors},}\ }\href {\doibase
  10.1103/PhysRevB.93.201116} {\bibfield  {journal} {\bibinfo  {journal} {Phys.
  Rev. B}\ }\textbf {\bibinfo {volume} {93}},\ \bibinfo {pages} {201116}
  (\bibinfo {year} {2016})}\BibitemShut {NoStop}%
\bibitem [{\citenamefont {Thomale}\ \emph {et~al.}(2013)\citenamefont
  {Thomale}, \citenamefont {Rachel},\ and\ \citenamefont
  {Schmitteckert}}]{thomale}%
  \BibitemOpen
  \bibfield  {author} {\bibinfo {author} {\bibfnamefont {Ronny}\ \bibnamefont
  {Thomale}}, \bibinfo {author} {\bibfnamefont {Stephan}\ \bibnamefont
  {Rachel}}, \ and\ \bibinfo {author} {\bibfnamefont {Peter}\ \bibnamefont
  {Schmitteckert}},\ }\bibfield  {title} {\enquote {\bibinfo {title} {Tunneling
  spectra simulation of interacting {M}ajorana wires},}\ }\href {\doibase
  10.1103/PhysRevB.88.161103} {\bibfield  {journal} {\bibinfo  {journal} {Phys.
  Rev. B}\ }\textbf {\bibinfo {volume} {88}},\ \bibinfo {pages} {161103}
  (\bibinfo {year} {2013})}\BibitemShut {NoStop}%
\bibitem [{\citenamefont {Chen}\ \emph {et~al.}(2011)\citenamefont {Chen},
  \citenamefont {Gu},\ and\ \citenamefont {Wen}}]{class2}%
  \BibitemOpen
  \bibfield  {author} {\bibinfo {author} {\bibfnamefont {Xie}\ \bibnamefont
  {Chen}}, \bibinfo {author} {\bibfnamefont {Zheng-Cheng}\ \bibnamefont {Gu}},
  \ and\ \bibinfo {author} {\bibfnamefont {Xiao-Gang}\ \bibnamefont {Wen}},\
  }\bibfield  {title} {\enquote {\bibinfo {title} {Classification of gapped
  symmetric phases in one-dimensional spin systems},}\ }\href {\doibase
  10.1103/PhysRevB.83.035107} {\bibfield  {journal} {\bibinfo  {journal} {Phys.
  Rev. B}\ }\textbf {\bibinfo {volume} {83}},\ \bibinfo {pages} {035107}
  (\bibinfo {year} {2011})}\BibitemShut {NoStop}%
\bibitem [{\citenamefont {Fidkowski}\ and\ \citenamefont
  {Kitaev}(2010)}]{class1}%
  \BibitemOpen
  \bibfield  {author} {\bibinfo {author} {\bibfnamefont {Lukasz}\ \bibnamefont
  {Fidkowski}}\ and\ \bibinfo {author} {\bibfnamefont {Alexei}\ \bibnamefont
  {Kitaev}},\ }\bibfield  {title} {\enquote {\bibinfo {title} {Effects of
  interactions on the topological classification of free fermion systems},}\
  }\href {\doibase 10.1103/PhysRevB.81.134509} {\bibfield  {journal} {\bibinfo
  {journal} {Phys. Rev. B}\ }\textbf {\bibinfo {volume} {81}},\ \bibinfo
  {pages} {134509} (\bibinfo {year} {2010})}\BibitemShut {NoStop}%
\bibitem [{\citenamefont {Katsura}\ \emph {et~al.}(2015)\citenamefont
  {Katsura}, \citenamefont {Schuricht},\ and\ \citenamefont
  {Takahashi}}]{Katsura}%
  \BibitemOpen
  \bibfield  {author} {\bibinfo {author} {\bibfnamefont {Hosho}\ \bibnamefont
  {Katsura}}, \bibinfo {author} {\bibfnamefont {Dirk}\ \bibnamefont
  {Schuricht}}, \ and\ \bibinfo {author} {\bibfnamefont {Masahiro}\
  \bibnamefont {Takahashi}},\ }\bibfield  {title} {\enquote {\bibinfo {title}
  {Exact ground states and topological order in interacting {K}itaev/{M}ajorana
  chains},}\ }\href {\doibase 10.1103/PhysRevB.92.115137} {\bibfield  {journal}
  {\bibinfo  {journal} {Phys. Rev. B}\ }\textbf {\bibinfo {volume} {92}},\
  \bibinfo {pages} {115137} (\bibinfo {year} {2015})}\BibitemShut {NoStop}%
\bibitem [{\citenamefont {Alicea}\ and\ \citenamefont
  {Fendley}(2016)}]{StrongZeroMode1}%
  \BibitemOpen
  \bibfield  {author} {\bibinfo {author} {\bibfnamefont {Jason}\ \bibnamefont
  {Alicea}}\ and\ \bibinfo {author} {\bibfnamefont {Paul}\ \bibnamefont
  {Fendley}},\ }\bibfield  {title} {\enquote {\bibinfo {title} {Topological
  phases with parafermions: Theory and blueprints},}\ }\href {\doibase
  10.1146/annurev-conmatphys-031115-011336} {\bibfield  {journal} {\bibinfo
  {journal} {Annu. Rev. of Condens. Matter Phys.}\ }\textbf {\bibinfo {volume}
  {7}},\ \bibinfo {pages} {119--139} (\bibinfo {year} {2016})}\BibitemShut
  {NoStop}%
\bibitem [{\citenamefont {Else}\ \emph {et~al.}(2017)\citenamefont {Else},
  \citenamefont {Fendley}, \citenamefont {Kemp},\ and\ \citenamefont
  {Nayak}}]{prethermal}%
  \BibitemOpen
  \bibfield  {author} {\bibinfo {author} {\bibfnamefont {Dominic~V.}\
  \bibnamefont {Else}}, \bibinfo {author} {\bibfnamefont {Paul}\ \bibnamefont
  {Fendley}}, \bibinfo {author} {\bibfnamefont {Jack}\ \bibnamefont {Kemp}}, \
  and\ \bibinfo {author} {\bibfnamefont {Chetan}\ \bibnamefont {Nayak}},\
  }\bibfield  {title} {\enquote {\bibinfo {title} {Prethermal strong zero modes
  and topological qubits},}\ }\href {\doibase 10.1103/PhysRevX.7.041062}
  {\bibfield  {journal} {\bibinfo  {journal} {Phys. Rev. X}\ }\textbf {\bibinfo
  {volume} {7}},\ \bibinfo {pages} {041062} (\bibinfo {year}
  {2017})}\BibitemShut {NoStop}%
\bibitem [{\citenamefont {Goldstein}\ and\ \citenamefont
  {Chamon}(2012)}]{Goldstein}%
  \BibitemOpen
  \bibfield  {author} {\bibinfo {author} {\bibfnamefont {G.}~\bibnamefont
  {Goldstein}}\ and\ \bibinfo {author} {\bibfnamefont {C.}~\bibnamefont
  {Chamon}},\ }\bibfield  {title} {\enquote {\bibinfo {title} {Exact zero modes
  in closed systems of interacting fermions},}\ }\href {\doibase
  10.1103/PhysRevB.86.115122} {\bibfield  {journal} {\bibinfo  {journal} {Phys.
  Rev. B}\ }\textbf {\bibinfo {volume} {86}},\ \bibinfo {pages} {115122}
  (\bibinfo {year} {2012})}\BibitemShut {NoStop}%
\bibitem [{\citenamefont {Kemp}\ \emph {et~al.}(2017)\citenamefont {Kemp},
  \citenamefont {Yao}, \citenamefont {Laumann},\ and\ \citenamefont
  {Fendley}}]{Kemp2017}%
  \BibitemOpen
  \bibfield  {author} {\bibinfo {author} {\bibfnamefont {Jack}\ \bibnamefont
  {Kemp}}, \bibinfo {author} {\bibfnamefont {Norman~Y}\ \bibnamefont {Yao}},
  \bibinfo {author} {\bibfnamefont {Christopher~R}\ \bibnamefont {Laumann}}, \
  and\ \bibinfo {author} {\bibfnamefont {Paul}\ \bibnamefont {Fendley}},\
  }\bibfield  {title} {\enquote {\bibinfo {title} {Long coherence times for
  edge spins},}\ }\href {https://doi.org/10.1088/1742-5468/aa73f0} {\bibfield
  {journal} {\bibinfo  {journal} {J. Stat. Mech.}\ }\textbf {\bibinfo {volume}
  {2017}},\ \bibinfo {pages} {P063105} (\bibinfo {year} {2017})}\BibitemShut
  {NoStop}%
\bibitem [{mul()}]{multigamma}%
  \BibitemOpen
  \href@noop {} {}\bibinfo {note} {Also, within the proposed approach, terms
  including three or more $\gamma$'s can (and in a general case should) be
  taken into account. We have checked at least for at least a few paprameter
  sets that the contribution from terms with three $\gamma$'s is
  negligible.}\BibitemShut {Stop}%
\bibitem [{sup()}]{supp}%
  \BibitemOpen
  \href@noop {} {}\bibinfo {note} {See Supplemental Material at [URL will be
  inserted by publisher] for the details of the finite size scaling, results
  for nearest-neighbor interaction, and more formal discussion of the Majorana
  zero modes.}\BibitemShut {Stop}%
\bibitem [{\citenamefont {Fendley}(2016)}]{StrongZeroMode0}%
  \BibitemOpen
  \bibfield  {author} {\bibinfo {author} {\bibfnamefont {Paul}\ \bibnamefont
  {Fendley}},\ }\bibfield  {title} {\enquote {\bibinfo {title} {Strong zero
  modes and eigenstate phase transitions in the {XYZ}/interacting {M}ajorana
  chain},}\ }\href {http://stacks.iop.org/1751-8121/49/i=30/a=30LT01}
  {\bibfield  {journal} {\bibinfo  {journal} {J. Phys. A}\ }\textbf {\bibinfo
  {volume} {49}},\ \bibinfo {pages} {30LT01} (\bibinfo {year}
  {2016})}\BibitemShut {NoStop}%
\bibitem [{\citenamefont {Jermyn}\ \emph {et~al.}(2014)\citenamefont {Jermyn},
  \citenamefont {Mong}, \citenamefont {Alicea},\ and\ \citenamefont
  {Fendley}}]{StrongZeroMode2}%
  \BibitemOpen
  \bibfield  {author} {\bibinfo {author} {\bibfnamefont {Adam~S.}\ \bibnamefont
  {Jermyn}}, \bibinfo {author} {\bibfnamefont {Roger S.~K.}\ \bibnamefont
  {Mong}}, \bibinfo {author} {\bibfnamefont {Jason}\ \bibnamefont {Alicea}}, \
  and\ \bibinfo {author} {\bibfnamefont {Paul}\ \bibnamefont {Fendley}},\
  }\bibfield  {title} {\enquote {\bibinfo {title} {Stability of zero modes in
  parafermion chains},}\ }\href {\doibase 10.1103/PhysRevB.90.165106}
  {\bibfield  {journal} {\bibinfo  {journal} {Phys. Rev. B}\ }\textbf {\bibinfo
  {volume} {90}},\ \bibinfo {pages} {165106} (\bibinfo {year}
  {2014})}\BibitemShut {NoStop}%
\bibitem [{\citenamefont {Mierzejewski}\ \emph
  {et~al.}(2015{\natexlab{a}})\citenamefont {Mierzejewski}, \citenamefont
  {Prelov\ifmmode~\check{s}\else \v{s}\fi{}ek},\ and\ \citenamefont
  {Prosen}}]{mm1}%
  \BibitemOpen
  \bibfield  {author} {\bibinfo {author} {\bibfnamefont {Marcin}\ \bibnamefont
  {Mierzejewski}}, \bibinfo {author} {\bibfnamefont {Peter}\ \bibnamefont
  {Prelov\ifmmode~\check{s}\else \v{s}\fi{}ek}}, \ and\ \bibinfo {author}
  {\bibfnamefont {Toma{\v{z}}}\ \bibnamefont {Prosen}},\ }\bibfield  {title}
  {\enquote {\bibinfo {title} {Identifying local and quasilocal conserved
  quantities in integrable systems},}\ }\href {\doibase
  10.1103/PhysRevLett.114.140601} {\bibfield  {journal} {\bibinfo  {journal}
  {Phys. Rev. Lett.}\ }\textbf {\bibinfo {volume} {114}},\ \bibinfo {pages}
  {140601} (\bibinfo {year} {2015}{\natexlab{a}})}\BibitemShut {NoStop}%
\bibitem [{\citenamefont {O'Brien}\ and\ \citenamefont
  {Wright}()}]{OBrien:2015aks}%
  \BibitemOpen
  \bibfield  {author} {\bibinfo {author} {\bibfnamefont {T.~E.}\ \bibnamefont
  {O'Brien}}\ and\ \bibinfo {author} {\bibfnamefont {A.~R.}\ \bibnamefont
  {Wright}},\ }\bibfield  {title} {\enquote {\bibinfo {title} {{A many-body
  interpretation of {M}ajorana bound states, and conditions for their
  localisation}},}\ }\href@noop {} {\ }\Eprint
  {http://arxiv.org/abs/1508.06638} {arXiv:1508.06638} \BibitemShut {NoStop}%
\bibitem [{\citenamefont {Rigol}\ and\ \citenamefont
  {Shastry}(2008)}]{shastry}%
  \BibitemOpen
  \bibfield  {author} {\bibinfo {author} {\bibfnamefont {Marcos}\ \bibnamefont
  {Rigol}}\ and\ \bibinfo {author} {\bibfnamefont {B.~S.}\ \bibnamefont
  {Shastry}},\ }\bibfield  {title} {\enquote {\bibinfo {title} {Drude weight in
  systems with open boundary conditions},}\ }\href {\doibase
  10.1103/PhysRevB.77.161101} {\bibfield  {journal} {\bibinfo  {journal} {Phys.
  Rev. B}\ }\textbf {\bibinfo {volume} {77}},\ \bibinfo {pages} {161101}
  (\bibinfo {year} {2008})}\BibitemShut {NoStop}%
\bibitem [{\citenamefont {Sirker}\ \emph {et~al.}(2014)\citenamefont {Sirker},
  \citenamefont {Konstantinidis}, \citenamefont {Andraschko},\ and\
  \citenamefont {Sedlmayr}}]{sirker2014}%
  \BibitemOpen
  \bibfield  {author} {\bibinfo {author} {\bibfnamefont {J.}~\bibnamefont
  {Sirker}}, \bibinfo {author} {\bibfnamefont {N.~P.}\ \bibnamefont
  {Konstantinidis}}, \bibinfo {author} {\bibfnamefont {F.}~\bibnamefont
  {Andraschko}}, \ and\ \bibinfo {author} {\bibfnamefont {N.}~\bibnamefont
  {Sedlmayr}},\ }\bibfield  {title} {\enquote {\bibinfo {title} {Locality and
  thermalization in closed quantum systems},}\ }\href {\doibase
  10.1103/PhysRevA.89.042104} {\bibfield  {journal} {\bibinfo  {journal} {Phys.
  Rev. A}\ }\textbf {\bibinfo {volume} {89}},\ \bibinfo {pages} {042104}
  (\bibinfo {year} {2014})}\BibitemShut {NoStop}%
\bibitem [{\citenamefont {Mierzejewski}\ \emph
  {et~al.}(2015{\natexlab{b}})\citenamefont {Mierzejewski}, \citenamefont
  {Prosen},\ and\ \citenamefont {Prelov\ifmmode~\check{s}\else
  \v{s}\fi{}ek}}]{mm2}%
  \BibitemOpen
  \bibfield  {author} {\bibinfo {author} {\bibfnamefont {Marcin}\ \bibnamefont
  {Mierzejewski}}, \bibinfo {author} {\bibfnamefont {Toma{\v{z}}}\ \bibnamefont
  {Prosen}}, \ and\ \bibinfo {author} {\bibfnamefont {Peter}\ \bibnamefont
  {Prelov\ifmmode~\check{s}\else \v{s}\fi{}ek}},\ }\bibfield  {title} {\enquote
  {\bibinfo {title} {Approximate conservation laws in perturbed integrable
  lattice models},}\ }\href {\doibase 10.1103/PhysRevB.92.195121} {\bibfield
  {journal} {\bibinfo  {journal} {Phys. Rev. B}\ }\textbf {\bibinfo {volume}
  {92}},\ \bibinfo {pages} {195121} (\bibinfo {year}
  {2015}{\natexlab{b}})}\BibitemShut {NoStop}%
\bibitem [{\citenamefont {Prelov{\v{s}}ek}\ \emph {et~al.}(2017)\citenamefont
  {Prelov{\v{s}}ek}, \citenamefont {Mierzejewski}, \citenamefont
  {Bari{\v{s}}i{\v{c}}},\ and\ \citenamefont {Herbrych}}]{annalen}%
  \BibitemOpen
  \bibfield  {author} {\bibinfo {author} {\bibfnamefont {P.}~\bibnamefont
  {Prelov{\v{s}}ek}}, \bibinfo {author} {\bibfnamefont {M.}~\bibnamefont
  {Mierzejewski}}, \bibinfo {author} {\bibfnamefont {O.}~\bibnamefont
  {Bari{\v{s}}i{\v{c}}}}, \ and\ \bibinfo {author} {\bibfnamefont
  {J.}~\bibnamefont {Herbrych}},\ }\bibfield  {title} {\enquote {\bibinfo
  {title} {Density correlations and transport in models of many-body
  localization},}\ }\href {http:https://doi.org/10.1002/andp.201600362}
  {\bibfield  {journal} {\bibinfo  {journal} {Ann. Phys. (Amsterdam)}\ }\textbf
  {\bibinfo {volume} {529}} (\bibinfo {year} {2017})}\BibitemShut {NoStop}%
\bibitem [{\citenamefont {Miao}\ \emph {et~al.}(2017)\citenamefont {Miao},
  \citenamefont {Jin}, \citenamefont {Zhang},\ and\ \citenamefont
  {Zhou}}]{Miao2017}%
  \BibitemOpen
  \bibfield  {author} {\bibinfo {author} {\bibfnamefont {Jian-Jian}\
  \bibnamefont {Miao}}, \bibinfo {author} {\bibfnamefont {Hui-Ke}\ \bibnamefont
  {Jin}}, \bibinfo {author} {\bibfnamefont {Fu-Chun}\ \bibnamefont {Zhang}}, \
  and\ \bibinfo {author} {\bibfnamefont {Yi}~\bibnamefont {Zhou}},\ }\bibfield
  {title} {\enquote {\bibinfo {title} {Exact solution for the interacting
  {K}itaev chain at the symmetric point},}\ }\href {\doibase
  10.1103/PhysRevLett.118.267701} {\bibfield  {journal} {\bibinfo  {journal}
  {Phys. Rev. Lett.}\ }\textbf {\bibinfo {volume} {118}},\ \bibinfo {pages}
  {267701} (\bibinfo {year} {2017})}\BibitemShut {NoStop}%
\end{thebibliography}%

\end{document}